# Perceptual Sensitivity to Stereo Geometry Errors in Head-Mounted Displays

RAFFLES XINGQI ZHU, Reality Labs Research, Meta, USA McGill University, Canada
CHARLIE S. BURLINGHAM, Reality Labs Research, Meta, USA
OLIVIER MERCIER, Reality Labs Research, Meta, USA
PHILLIP GUAN, Reality Labs Research, Meta, USA

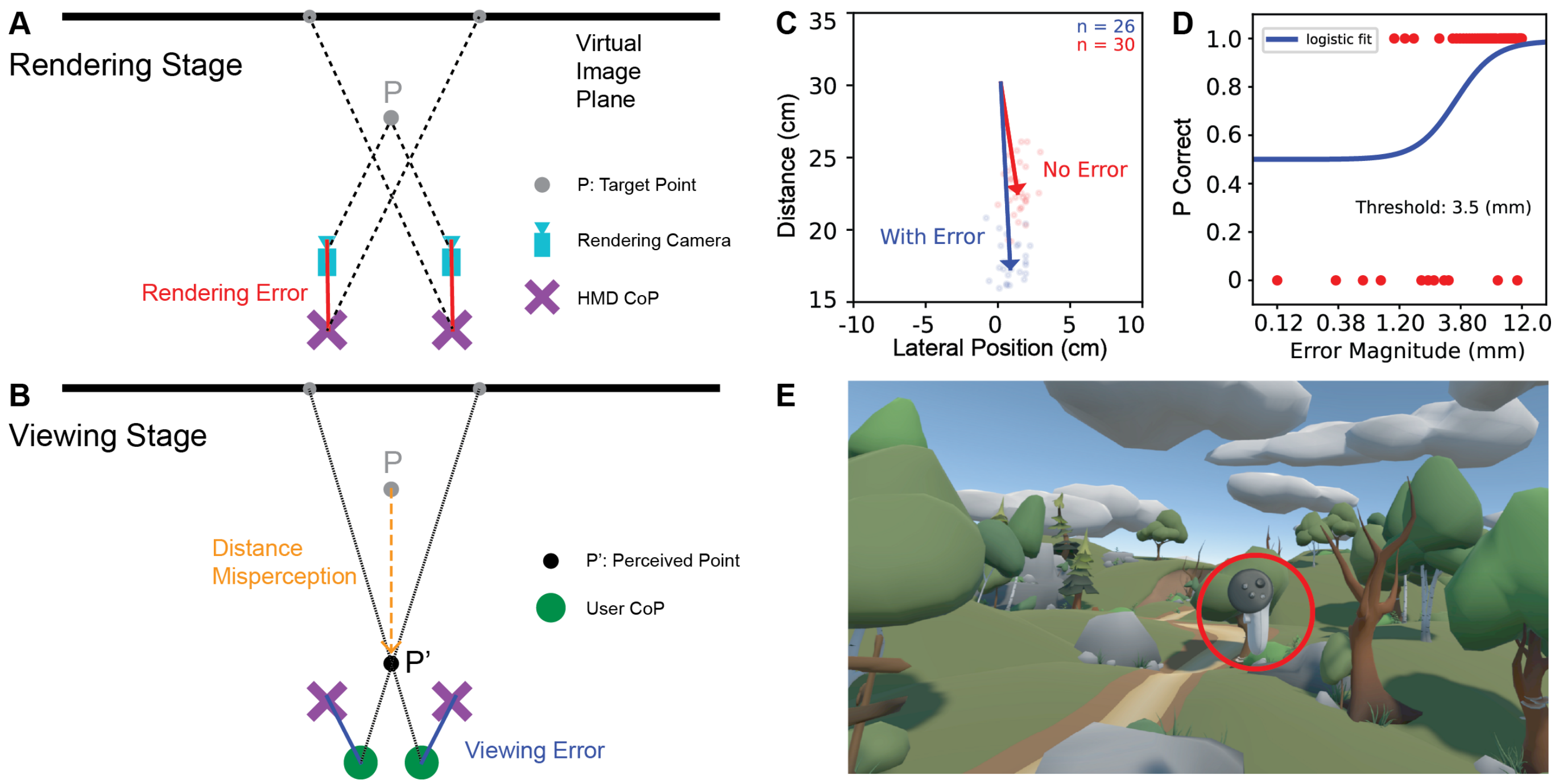

Fig. 1. We investigate whether geometric predictions from purposefully introduced rendering and viewing stereo geometry errors in head-mounted displays (HMDs) accurately model changes in perceived distance when these errors are viewed in an HMD. Some of the errors we examined would traditionally require physically changing the position of an HMD on the user's head (e.g., to simulate changes in eye relief). Precise manipulations of headset fit are difficult in practice, so we built an HMD software platform that allowed us render images in our HMD consistent with making these physical manipulations to facilitate our user studies. The impact of these errors was measured in a series of experiments involving blind reaching and visual only distance discrimination. **(A)** Rendering errors are displacements of the rendering cameras from the HMD centers of projection (CoP), typically located at the center of the eyebox. In this example, the rendering cameras are too far forward which can occur when video passthrough images are directly streamed to the HMD displays without view reprojection. Here the ray angle from the camera to a target point P is captured by the camera, and this ray is projected onto the display from the HMD CoP leading to the incorrect render geometry (the correct ray should pass from the HMD CoP through the target point P to the display). **(B)** Independent from rendering errors, viewing errors are displacements of the CoP of the viewer's eyes (user CoP) from the HMD CoP. Here, the incorrectly rendered left and right stereo images from (A) are additionally viewed from an incorrect location. The HMD interaxial distance (IAD, the distance between HMD CoP) is too large relative to the viewer's interpupillary distance (IPD). The viewer's eye relief is also too large, causing their eyes to be behind the HMD CoP. A simple ray-intersection model predicts that the combined effects of rendering and viewing errors will result in a misperception of the black point P' from the intended gray point P. **(C)** We evaluated the efficacy of this triangulation model in a blind reaching experiment, which required us to first use an eye-tracked HMD to compensate for naturally occurring viewing errors so that we could properly evaluate simulated rendering and viewing errors. This plot shows the blind reach performance for one participant in the no rendering or viewing error condition (red) and when HMD IAD is higher than user IPD (blue). In the absence of error, we find that there is still an under-reaching bias. Removing this bias aligns blind reach distances with predictions of our triangulation model for two out of three error conditions tested. **(D)** We measured error detection thresholds using a two interval forced choice psychophysical task. This design was made possible by leveraging eye tracking in our HMD platform to compare no-error stereo geometry to perturbed geometry due to introduced rendering and viewing errors. One example participant's binary response data for IPD viewing error are shown in red, with the psychometric function shown in blue. **(E)** The scene used in our experiments (controller in the reaching experiments is highlighted for emphasis).

Virtual and augmented reality head-mounted displays (HMDs) render and display head-tracked, stereo images to create an egocentric, 3D percept to the viewer. Errors in stereo geometry introduced by rendering from an incorrect location (e.g., from tracking errors) or viewing from the wrong position (e.g., from fitment errors) can alter the depth and distance a user perceives compared to the intended geometry. In this work we evaluate how accurately a purely geometric framework that models distance misperception arising from inaccurate stereo geometry can predict changes to user reach behavior. We show that some stereo geometry errors can induce both under- and over-reaching, and the magnitude of error is consistent with geometric predictions. Surprisingly, we find the mismatches between HMD display separation and viewer interpupillary distance error have significantly less impact on reach behavior than predicted by geometry prediction. In a follow-up psychophysical study we find a relatively large detection threshold (≈6 mm) to mismatches between viewer interpupillary distance and headset display separation, confirming the reaching result. We additionally show how perceptual thresholds to various stereo geometry errors change with viewing distance and scene complexity.

## 1 INTRODUCTION

Precise interaction in head-mounted displays (HMDs) requires correct stereo geometry for users to accurately perceive size and distance in virtual environments. Precisely reproducing this geometry requires co-location of three features: for each eye, the Render Camera Center of Projection (CoP), the HMD Display CoP, and the User's Eye CoP must be spatially aligned. In practice, this is rarely achieved and many geometric models have been built by researchers to predict the perceptual consequences that may result from these errors [Held and Banks 2008; Krajancich et al. 2020; Robinett and Rolland 1992; Woods et al. 1993].

The stereo-projection pipeline is susceptible to two classes of errors. First, rendering errors arise when the render camera used to generate the scene is displaced from the HMD CoP. Rendering errors are commonly introduced due to inaccuracies in tracking the user's head position [Holloway 1997] and from motion-to-photon latency [Allison et al. 1999]. In more recently introduced mixed-reality HMDs with video passthrough, rendering errors can be introduced by the static displacement of the see-through video camera in front of user's eyes [Biocca and Rolland 1998; Kuo et al. 2023; Lee et al. 2013].

Second, viewing errors occur when the rendered image is geometrically correct for a nominal eye position, but the user's actual eye (more specifically, the CoP of the eye) is displaced from this location [Banks et al. 2014; Guan et al. 2023; Held and Banks 2008; Vishwanath et al. 2005]. Current HMD optical designs assume a fixed, nominal eye position within the "eyebox" to maximize field of view and minimize distortion. HMDs without eyetracking assume the eye is always at this nominal location [Rong 2025]. Yet, variance in human facial structure, improper headset fit, or mismatches between the headset's inter-axial distance (IAD) and the user's interpupillary distance (IPD) inevitably place the user's Eye CoP away from the Display CoP. Even dynamic factors, such as the movement of the Eye CoP during eye rotations (i.e., ocular parallax), introduce transient viewing errors that most current rendering pipelines ignore [Holloway 1997; Rolland et al. 2004; Wann et al. 1995]. These displacements fundamentally alter the triangulation geometry of the viewed image that should, according to projective geometry, disrupt the viewer's perception of size and distance.

While the theoretical distortions resulting from these rendering and viewing errors can be modeled using stereo geometry, the predictive accuracy of a simple triangulation model remains an open question. Can human perception be predicted simply from first-order geometry, or does the visual system employ higher-order perceptual mechanisms to normalize the percept? Assessing the predictive power of these geometric models is difficult, as manipulating parameters like headset fit, tracking accuracy, and latency are often difficult or infeasible to readily adjust in real headsets.

In this work, we evaluate whether a simple geometric model can predict perceptual responses to errors in stereo geometry. To explore this, we make the following contributions:

- We build a software platform that allows direct manipulation of rendering and viewing errors in a real headset thereby enabling precise and repeatable adjustments of viewing errors without the need to physically manipulate headset fit. We used this platform to conduct a series of user studies and compared these results to the theoretical geometric model, and we additionally provide a web application that can be used to explore these geometric predictions interactively.
- We show that direct passthrough rendering errors and eye relief viewing errors induce changes in reach behavior that are well-predicted by stereo geometry, but we also identify a significant deviation in the case of IPD mismatch, which has a surprisingly small impact compared to model predictions.
- We conduct a psychophysical experiment exploring perceptual sensitivity to direct passthrough and IPD errors. We find that sensitivity varies with viewing distance and scene content, and confirm a surprising tolerance to IPD errors (≈ 6 mm), even at near distances.

## 2 RELATED WORK

*Distance Underestimation in HMDs.* Over the past several decades many studies have found that distances are underestimated in virtual reality (VR) [Creem-Regehr et al. 2015; El Jamiy and Marsh 2019a,b; Renner et al. 2013]. A variety of theories have been proposed to explain this phenomenon, but a comprehensive explanation has not been found [Creem-Regehr et al. 2023]. Proposed theories on the underlying cause of systematic distance underestimation in HMDs are typically related to effects of HMD hardware, such as field of view (FOV) [Buck et al. 2018; Kline and Witmer 1996], weight [Buck et al. 2018; Masnadi et al. 2022, 2021; Willemsen et al. 2009], display resolution [Kelly 2022; Ryu et al. 2005; Thompson et al. 2004], and higher-order perceptual effects like scene complexity, realism, presence, and embodiment [Bodenheimer et al. 2023; El Jamiy et al. 2020; Gonzalez-Franco et al. 2019; Vaziri et al. 2017]. In a meta analysis, Kelly [2022] found that distance is more accurately estimated in newer HMDs. Kelly explored the impact of technical characteristics like FOV and resolution on the magnitude of underestimation across studies, but there are other improvements in newer hardware that are more difficult to quantify such as more accurate head tracking, reduced distortion from improved optical designs, and better fitment systems. These advancements all improve the underlying stereo projection and relatively little work has investigated the effect low-level errors in stereo geometry may have on distance perception in HMDs. Our work is the first to explicitly measure how errors in stereo geometry affect behavior (reaching) which may offer new insights on topics like distance underestimation.

*Reaching Experiments in Personal Space.* There are fewer studies on distance estimation in VR for personal space (<2 meters) compared to action space (2-30 meters). Results are mixed showing overestimation [Rolland et al. 1995], underestimation [Napieralski et al. 2011], and accurate estimation [Naceri et al. 2011]. Similar to studies in action space, distance is typically measured using open-loop motor tasks such as blind reaching/pointing or verbal reports. Napieralski et. al [2011] found the difference between measurement protocal (blind reaching vs. verbal response) is much greater than the difference between VR and the real world (underestimation in both cases). Closed-loop visual feedback has been shown to improve distance judgement accuracy [Altenhoff et al. 2012; Kelly et al. 2014; Mohler et al. 2006], with visual motor re-calibration leading to more accurate reaching [Ebrahimi et al. 2014a, 2015b]. One challenge in characterizing reach distance in near space is co-registration of the viewer's egocentric coordinate frame and the coordinate frame of the system used to measure reach distance. Our stereo geometry framework in Section 3 explores how differences in eye relief in particular may introduce discrepancies between how far a user perceives their reach to be and physically measured reach distance by an external observer.

*Perceptual Effects of Inaccurate Geometry in HMDs.* Geometric errors have been shown to affect distance perception experimentally, but most studies evaluate engineering-related errors rather than geometric errors explicitly. For example, geometric calibration alleviates distance underestimation in AR [Kellner et al. 2012]. Magnification also causes significant changes in distance judgments [Kellner et al. 2012; Kuhl et al. 2009; Li et al. 2015; Zhang et al. 2012]. Some studies have examined the impact of mismatched HMD lens interaxial distance (IAD) to the user's IPD and have found that distance perception is minimally impacted by the IAD-IPD mismatch in action space [Chakraborty et al. 2024; Hibbard et al. 2020; Willemsen et al. 2008], likely a result of stereo cues being less accurate at farther distances [McCann et al. 2018]. Guan et al. [2023] detection thresholds to eye relief, IPD, and translational viewing errors, but they do not quantify the impact of these thresholds on user behavior or perception. Many computational models have been presented that examine the effects of viewing errors (i.e., displacement of the user CoP from the HMD CoP) [Held and Banks 2008; Rolland et al. 2004; Woods et al. 1993], but fewer have considered the joint effects of simultaneous viewing and rendering errors [Holloway 1997]. In this work, we complement prior geometric modeling by also pairing geometric predictions with empirical user studies.

# 3 STEREO GEOMETRY AND EMULATION IN VR

This section outlines the stereo intersection geometry governing rendering and viewing errors. While the principles behind this model are individually straightforward, their interactions are complex. We encourage readers to explore these interactions in depth using the interactive visualizer provided in the supplementary materials (Figure 8). Additionally, we describe the error emulation platform used for the user studies presented in Sections 4 and 5.

## 3.1 Model Overview

The geometric consequences of rendering and viewing errors are determined by basic ray-intersection and have been derived in previous works [Held and Banks 2008; Holloway 1997; Rolland et al. 2004; Woods et al. 1993]. Rather than re-deriving these equations, we focus on the intuition regarding how rendering errors, viewing errors, virtual image distance (VID), and scene geometry interact. More details, and an interactive web application (Figure 8) can be found in the supplementary materials.

In a typical HMD pipeline, accurate stereo perception requires the render camera and the user's eye Center of Projection (CoP) to coincide with the HMD CoP. Deviations from this alignment can introduce errors at two stages:

(1) Rendering Errors: Displacements of the render camera relative to the HMD CoP (Figure 1A).
(2) Viewing Errors: Displacements of the user eye CoP relative to the HMD CoP (Figure 1B).

Figure 2A illustrates the ideal process: a sphere in the world is projected as a circle on the display, which is then projected to a circle on the user's retina. Subsequent panels in Figure 2 demonstrate the geometric consequences when the render cameras or user eyes are displaced from the HMD CoP. To contextualize these consequences, we next explore how misalignments between the viewer's egocentric coordinate frame and the headset's coordinate frame (e.g., via eye relief errors) can affect perceived distance.

## 3.2 Coordinate Systems and Distance Measurement

Distance perception in HMDs involves two distinct coordinate systems. The origin of the *HMD coordinate system* is centered between the two HMD CoPs; the intended virtual scene is defined within this frame. However, the user perceives distance relative to their own body. The *egocentric coordinate system* is therefore defined relative to the viewer, with the origin centered between the user's actual left and right eye CoPs.

Ideally, these two systems align. However, viewing errors—such as eye relief shifts—introduce offsets between the egocentric and HMD frames (this is a simplified framework that does not account for the optical effects of HMD optics). This misalignment has critical implications for measuring distance perception. For example, consider a viewer with a +3 cm eye relief error fixating on a target at the virtual image distance (VID) (Figure 9C). While the target's position is unchanged in the HMD frame, the user's eyes are physically closer to the VID. Consequently, the distance perceived in the egocentric frame is 3 cm shorter than the distance defined in the HMD frame (Figure 8).

Crucially, standard blind reaching tasks tracked via external systems typically report positions in HMD coordinates, thereby missing this egocentric shift. To address this, Section 4 leverages eye tracking to record reaching data that explicitly accounts for the offset between egocentric and HMD coordinates. This allows us to analyze data in either frame, distinguishing between reach accuracy (HMD frame) and accurately perceived distance (egocentric frame).

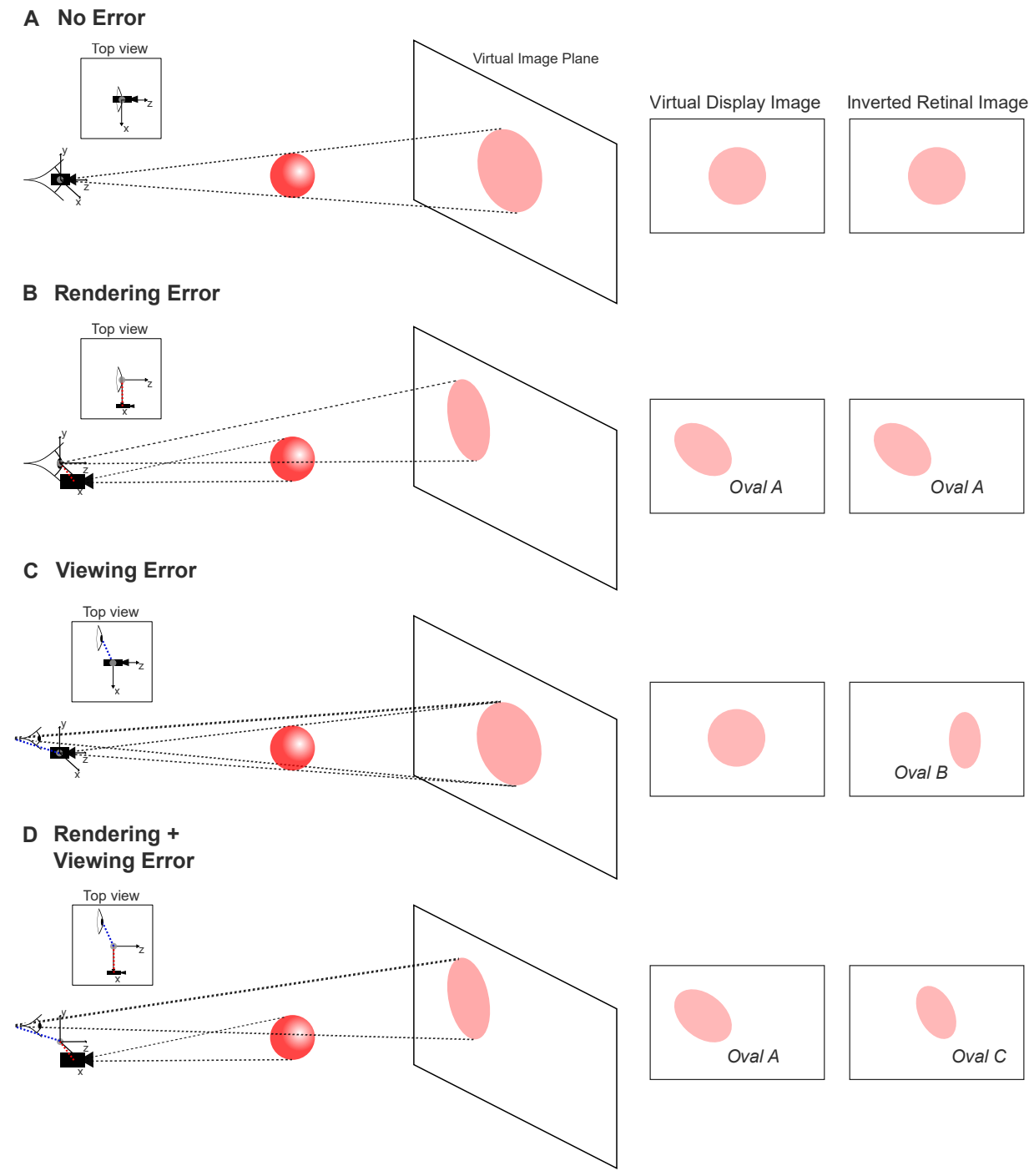

Fig. 2. Comparing rendering error with viewing error. For simplicity, we show a monocular example as the set up is the same for both eyes. The origin of the coordinate system is at the HMD CoP (grey dot). The virtual sphere is positioned directly in front of the the HMD CoP. **(A)** In the ideal scenario, the render camera, CoP of the user's eye, and HMD CoP are all at the same position in space. No rendering or viewing error exists in this system. The camera is on-axis with the sphere and will render the corret perspective projection (a circle) that will be presented on virtual image plane. The user will view this circle from the intended CoP and the user's retinal image of what is shown on the display plane is also a circle. **(B)** The render camera is displaced to the right and down from the HMD CoP, creating a rendering error. The camera will render the sphere from this incorrect location, and render an oval instead of a circle. This oval is shown on the display plane and the user views it from the correct position so they perceive the same incorrectly rendered oval. **(C)** The CoP of the user's eye is displaced to the left and behind the HMD CoP, creating a viewing error. Despite the rendered image being correct on the virtual image plane, the perceived image is distorted. **(D)** In realistic usage, rendering and viewing errors typically occur together and can result in geometric errors being introduced in both render and viewing stages.

### 3.3 HMD Error Simulation Platform

*Hardware.* We integrated a 240 Hz Tobii (Tobii AB, Sweden) eye tracker into a Meta Quest 3 (v72). For reaching experiments in Section 4, we tracked controller 3D position using the default Quest API. We measured controller tracking accuracy and found that tracking is accurate within 1.5 mm error across ten distances from 5 to 50 cm, measured from the front surface of the headset (see supplementary materials).

*Software.* We build a rendering pipeline in Unity (v2022.3.30f1) that can simulate the rendering and viewing errors outlined in Section 3.1. Our application uses multiple cameras and quads for ray capture and reprojection to accomplish this, and a high-level overview is provided here. Step-by-step rendering details are provided in the supplementary materials. First a pair of render cameras render the scene and sends their framebuffers to a pair of quads simulating the virtual displays in front of HMD CoP. The VID (virtual image distance) was set to 1.3 m, approximating a reasonable VID similar to most modern HMDs designed minimize vergence-accommodation conflict [Hoffman et al. 2008]. Render error can be applied as a displacement of the render camera from the HMD CoP. Next, a pair of "viewing" cameras (implemented as standard camera object) view the virtual display quads from the previous step. The viewing cameras are meant to simulate user CoP in our *simulated* headset space. Viewing error is applied as a displacement of the viewing camera from the HMD CoP. Lastly, the user's eyes are unlikely to be at the nominally assumed position within the Quest 3 eyebox, so we must also compensate for this naturally-occurring viewing error in the headset. This is accomplished by placing a pair of quads showing viewing camera's framebuffer in front of the actual user's 3D pupil positions from the eye tracker. These quads showing the user's expected view of the simulated rendering and/or viewing errors are imaged by the standard OVRCameras (which are positioned at actual headset CoP, not at the user's actual eye position) and this framebuffer is shown on the Quest 3 display.

## 4 HOW ACCURATELY DOES GEOMETRY PREDICT BLIND REACH DISTANCE?

To understand how stereo geometry errors in HMDs affect near-field distance perception, and specifically how changes in perceived distance impact reaching behavior, we compared participants' blind reaching errors against the stereo triangulation model described in in Section 3.1.

### 4.1 Error Condition Selection

We evaluated three rendering and viewing errors critical to current VR hardware designs (Figure 9).

*Direct Passthrough Rendering Error.* Video passthrough headsets can use view reprojection [El Chemaly et al. 2025; Xiao et al. 2022] or computational imaging [Kuo et al. 2023] to correct for the positional offset between the passthrough cameras and the viewer's eyes. Alternatively, "direct passthrough" avoids these computationally expensive methods, and potential image distortions associated with them, by directly streaming perspective-incorrect video. To simulate this effect, we displaced the render cameras by 5.5 cm—a value representative of the thickness of current commercial headsets. We also simulated a -5.5 cm error for completeness (Figure 9A).

*IPD Viewing Error.* Perceived depth is inherently tied to retinal binocular disparity and a user's IPD. Failure to match the headset's interaxial distance (IAD) to the user's IPD creates a viewing error (not a rendering error because the render cameras and HMD CoP are colocated) and the user will experience incorrect disparity and vergence demand at the fixation point. We picked 12 mm errors based on a worst-case error in real life HMD use (Figure 9B).

*Eye Relief Viewing Error.* Due to differences in facial geometry, a viewer's actual nominal eye position in the headset may be displaced from this assumed location. We picked 3 cm as a potential eye relief viewing error which represents a value that is as large as could be reasonably expected for a typical HMD optical design (Figure 9C).

## 4.2 Participants

Thirty-two participants (mean age: 32.2 years, $\sigma$ = 7.2; mean IPD: 62.7 mm, $\sigma$ = 2.6) with normal or corrected-to-normal visual acuity of 20/20 and Randot stereo acuity <= 50 arc seconds participated in the study. All study protocols were IRB approved.

## 4.3 Methods

*Real Life Blind Reaching.* To check for inherent reaching biases, participants viewed a physical coin at 30 cm, closed their eyes, and reached for it. During the task, each participant's head was stabilized in a chin rest. A coin was placed on the table at 30 cm away from their forehead by the experimenter and participants were asked to remember its position. They then closed their eyes, after which the coin was removed by the experimenter, and were asked to reach for the coin's location. Participant's eyes remained closed until reach distance was recorded and their arm was back at their side. The participant was not made aware that the coin was placed at the same distance on each trial. This process was repeated three times to assess baseline blind reaching performance in real life.

*In Headset Blind Reaching.* Participants viewed a virtual controller at 30 cm (Figure 1E), hid the controller via a button press, and reached for the remembered location with a real controller. They had unlimited time to view the virtual controller and pressed a button on the left controller to hide the target when they were ready to start their reach with the right controller. Once they were satisfied with their controller placement, participants pressed another button on the left controller to record the right controller's position, and returned their right hand with the controller to a comfortable resting position. There were seven conditions in total: ± 55 mm direct passthrough rendering error, ± 12 mm IPD viewing error, ± 30 mm eye relief error, and a no error condition. The testing order of conditions was random and participants completed a total of 210 trials (30 trials for each condition) in a single session. Participants were asked to minimize head movement. An outlier analysis was performed for each condition to remove trials that appeared to be inadvertent completions where at least one of three coordinates of the controller was recorded beyond lower or upper fences (1st quartile - 1.5 times the interquartile range (IQR) or 3rd quartile + 1.5 IQR) of the recorded reach positions. On average this led to the removal of 2 trails ($\sigma$ = 1.7) per condition (30 trials).

*Data Transforms.* Two transforms were applied to the reaching data. First we apply bias correction to account for each participant's baseline blind-reaching bias from the real-life reach (Figure 3, right marker). These transformed values are shown in Figure 4 (right marker of each panel). For the second transform, we converted distances from HMD coordinates to egocentric coordinates (Section 3.1) since perceived distance by definition is in an egocentric coordinate frame. The transformed egocentric distance values are shown in Figure 5. This second transform only affects values in the eye relief error condition as HMD coordinate frame and egocentric coordinate frame are offset by 3 cm (Figure 9C).

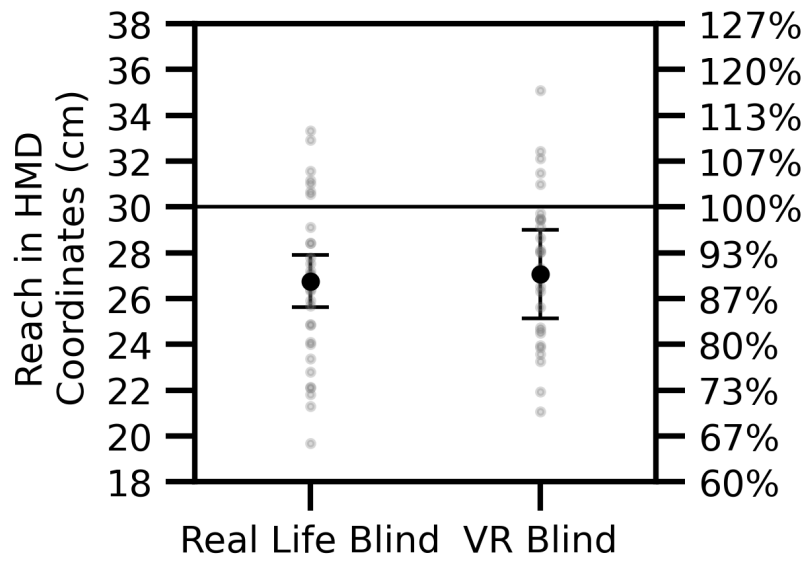

Fig. 3. Blind reaching with correct stereo geometry (no viewing or rendering errors). Black points show the average reach distance and 95% confidence intervals across all participants (n=32) for a target positioned 30 cm away in HMD coordinate system. Grey points show individual participant data. On average, participants under-reach in both real-life and VR blind reaching conditions by approximately 10%.

## 4.4 Statistics

Linear mixed effects models (LMEM) were used to estimate the influence of rendering and viewing errors on the magnitudes of reaching errors. In all cases, we specified a maximal model by default (i.e., including random intercepts and slopes for all relevant factors) following [Barr et al. 2013; Bates et al. 2015; Harrison et al. 2018; Maxwell et al. 2017]. For the reaching data, the maximal model was

$$\text{reachBias} \sim \text{error} + (\text{error} \mid \text{ID}) \tag{1}$$

where error was the magnitude in centimeters of the rendering or viewing error — for example, for direct passthrough: -5.5, 0, or 5.5 cm; reachBias was the bias of the average reach endpoint in egocentric coordinates, after also subtracting off any biases observed in the 0 cm / no-error condition. ID was simply each participant ID.

## 4.5 Results

Geometric errors induced statistically significant changes in perceived distance across all conditions. Notably, the magnitude of these shifts aligned closely with geometric predictions for direct passthrough and eye relief errors. In contrast, the observed effect of IPD error, while significant, was substantially smaller than the geometric model predicted.

*No Rendering or Viewing Error.* Participants consistently under-reached the 30 cm target (hypometric reaching) in both VR and real-world baselines (Figure 3). Across 32 participants, endpoints fell short by approximately 3–4 cm (~ 10%). The difference between real-world and VR performance was not statistically significant (paired t-test: $t = 0.71, p = 0.47, df = 31$), suggesting that the under-reaching observed in this study shares a common etiological origin

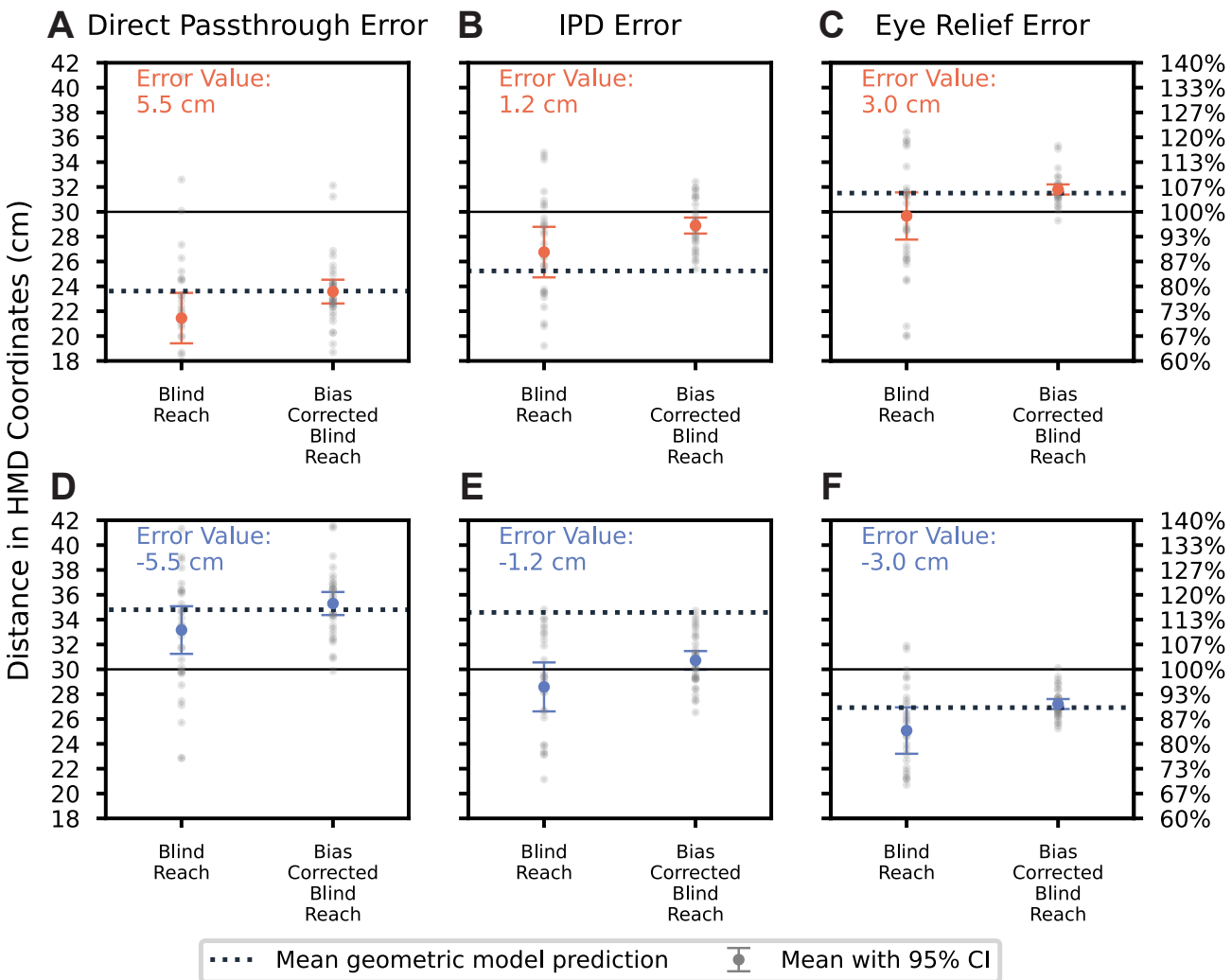

Fig. 4. Reach distance in HMD coordinates for blind reaching and blind reaching corrected for under-reaching bias for **(A)** direct passthrough error, **(B)** IPD error, and **(C)** eye relief error. Positive errors are shown in the top row (red) and negative errors are shown in the bottom row (blue). Black points are means with error bars denoting 95% confidence interval (n=32). After accounting for participant under-reaching bias, blind reach distances are well-predicted by our model for direct passthrough error and eye relief error.

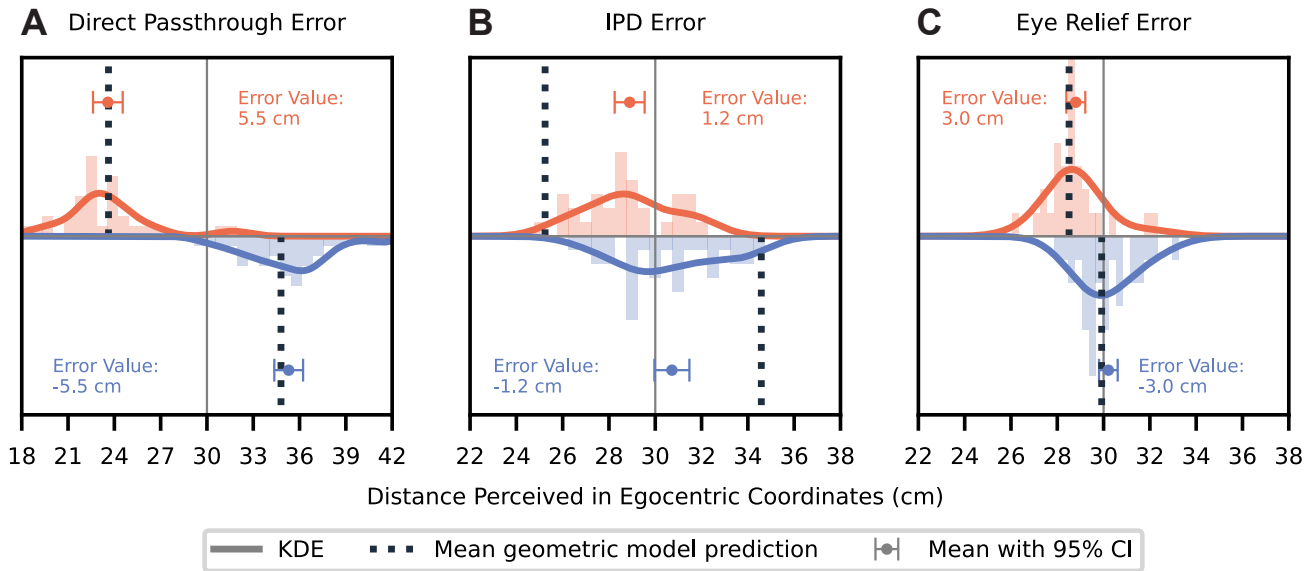

Fig. 5. Perceived distance in egocentric coordinates induced by rendering and viewing errors compared to model predictions. Each participant's blind reaching performance is adjusted to account for their no-error blind reach baseline for **(A)** direct passthrough error, **(B)** IPD error, and **(C)** eye relief error. For each, the red histogram shows individual data (n=32) for positive error while the yellow histogram is for negative error. A kernel density estimate (KDE) is shown to visualize the distribution. **(A)** and **(B)** replot data from Figure 4A-B (right most data points) since HMD coordinate system overlaps with egocentric coordinate system for those two error types. Our model predictions in egocentric coordinates provide good account of the data for direct passthrough error and eye relief error, but not for IPD error. A statistically significant effect is found in all three conditions when using a general linear mixed model to examine the relationship between error magnitudes and reach accuracy.

with natural proprioceptive or visual biases rather than being an artifact of the VR hardware.

*Direct Passthrough Error.* Participants underestimated distance when the render cameras were offset forward (Figure 5A). The LMEM revealed a near-perfect linear relationship between the physical offset and the perceptual error ($\beta = -1.07, t = -13.87, p < 0.0001$). This slope indicates a $\sim$ 1 cm underestimation of distance for every 1 cm of forward camera displacement. These results confirm that distance misperception caused by direct passthrough offsets is well-described by simple triangulation geometry.

*IPD Error.* Perceived distance scaled inversely with IPD error (Figure 5B). When the HMD IAD exceeded the user's IPD, targets appeared closer ($\beta = -0.76, t = -3.20, p < 0.02$). This implies a 0.76 cm underestimation for every 1 cm of excess IAD. While the effect is significant, the geometric model over-predicts the magnitude of this error, suggesting that the visual system may partially compensate for IPD mismatches (discussed further in Section 6).

*Eye Relief Error.* Eye relief errors produced a significant linear shift in perceived distance ($\beta = -0.23, t = -5.38, p < 0.0001$). A +3 cm error (eyes closer to display) caused the target to appear closer than intended. Conversely, a −3 cm error (eyes further from display) resulted in reaches that appeared accurate in the egocentric frame (Figure 5C).

This apparent accuracy is due to a cancellation effect at this specific target distance: the *optical error* (which geometrically pushes the virtual image away) is effectively neutralized by the *egocentric frame shift* (the user standing 3 cm further back). Although these factors canceled out at 30 cm, the linear model confirms that our geometric predictions accurately account for the underlying trend in both positive and negative directions.

## 5 HOW MUCH STEREO GEOMETRY ERROR IS DETECTABLE?

While direct passthrough errors elicited a reaching response consistent with geometric predictions, the impact of IPD errors was significantly weaker than predicted. This suggests that the visual system may be robust to—or perhaps less sensitive to—IPD mismatches compared to rendering offsets. To determine if this reduced behavioral sensitivity corresponds to a higher perceptual detection threshold, we conducted a follow-up psychophysical experiment using a two-interval forced choice (2IFC) task.

### 5.1 Participants

The same 32 participants that performed the reaching experiments participated in this study and all protocols were IRB approved.

### 5.2 Methods

In each trial, participants viewed a red sphere in two successive intervals: a reference interval (rendered with correct stereo geometry) and a comparison interval (rendered with either a direct passthrough or IPD error). The interval order was randomized, and each interval lasted 800 ms with a 200 ms inter-stimulus interval. Participants indicated which interval appeared closer via controller input. The monocular size of the sphere in the reference interval was always fixed at 1°, but scaled naturally according to perspective geometry in the comparison interval. We used the Method of Constant Stimuli to sample the psychometric function, varying error magnitudes in

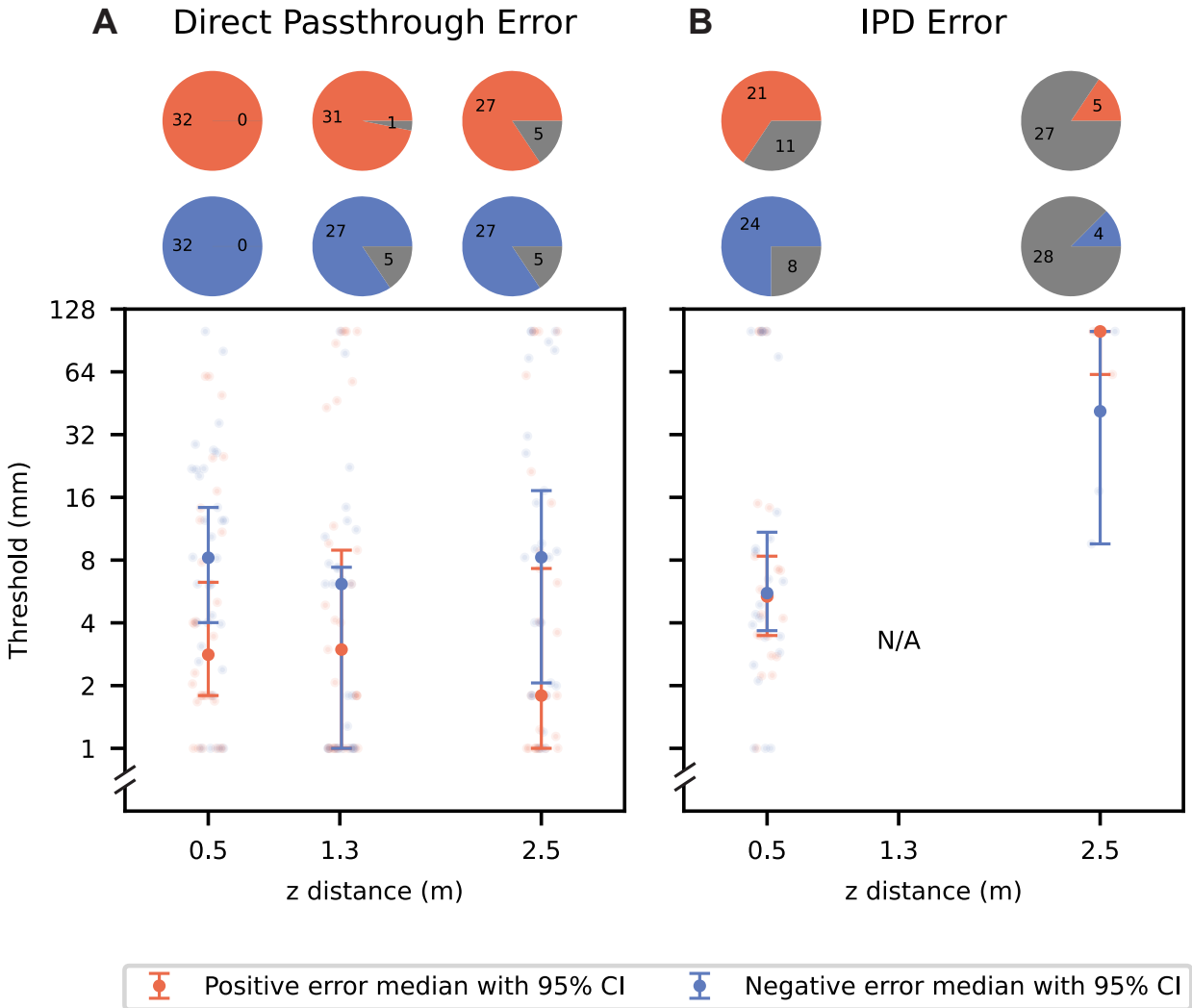

Fig. 6. Detection thresholds for **(A)** direct passthrough error and **(B)** IPD error. Red and blue points are detection thresholds for positive and negative error directions. For IPD error, a positive error means HMD IAD is larger than user IPD and vice versa. For direct passthrough error, a positive error means rendering cameras are in front of HMD CoP and vice versa. Pie chart shows split of participants that were sensitive to the specific error type (red or blue) and sign vs those who performed at chance (gray). For target placed 1.3 m and affected by IPD error, our geometric model predicts no error and detection threshold is ill-defined.

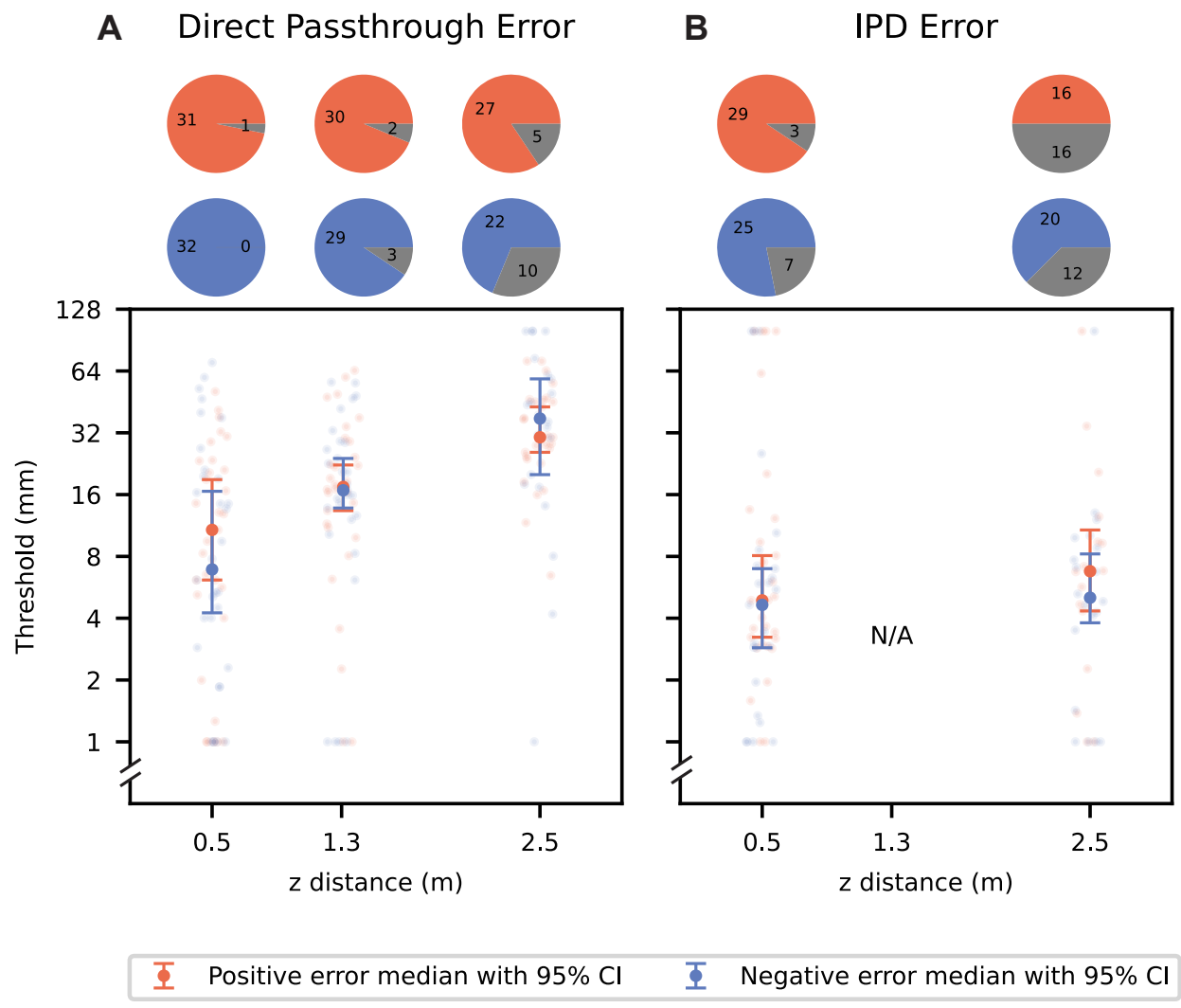

Fig. 7. Detection thresholds for **(A)** direct passthrough error and **(B)** IPD error with only the skybox scene.

100 equal steps (±12 mm for IPD; ±100 mm for direct passthrough). Each participant completed 1200 trials (2 error types × 3 distances × 2 scenes × 100 repetitions) in randomized order, separated into three blocks for each viewing distance.

## 5.3 Viewing Conditions

We expected sensitivity for both error types to depend on viewing distance and scene content. We varied distance and visual scene to explore the effects of three specific perceptual phenomenon.

*Stereo Sensitivity Gradient (IPD).* Binocular disparity cues degrade non-linearly with distance ($1/d^2$) [Howard and Rogers 2012]. We selected test distances of 0.5 m (Near), 1.3 m (Intermediate/VID), and 2.5 m (Far). We hypothesized that IPD errors, which rely on disparity and vergence shifts, would be more detectable in the near field, undetectable at the VID (because viewing errors do not introduce geometric errors at the VID), and less detectable 2.5 m.

*Weber's Law (Passthrough).* For direct passthrough, the rendering error represents a depth offset ($\Delta d$). According to Weber's Law, the Just Noticeable Difference (JND) should scale with total distance ($\Delta d/d = k$). Therefore, we expected thresholds to be lowest at 0.5 m and highest at 2.5 m.

*Cue Integration (Scene Content).* We utilized two environments to modulate cue reliability. A complex scene provided strong monocular cues (linear perspective, height-in-field) that may conflict with or override erroneous stereo cues. A sparse scene showed only a fixation target with a Skybox background, and this reduced scene may increase sensitivity to IPD errors compared to the structured environment.

## 5.4 Psychometric Function Fitting

We excluded participants who were not sensitive enough to the combination of error type and error sign by performing a one-tailed binomial test (alpha value 0.05), which determines whether psychometric performance is different from chance (50%). We estimated detection thresholds by fitting a psychometric function to the log-transformed error values for positive and negative errors separately. Additional details are provided in Supplementary Materials.

## 5.5 Results

*Error Detectability in Naturalistic Conditions.* We first assessed the absolute range of perceptible errors in the naturalistic environment (the Forest scene). Overall we identify trends that are consistent with our expectations, IPD error is perceptible at near distances and nearly imperceptible at far distances (Figure 6B). However, participants have a surprisingly high threshold for IPD errors, at approximately 6 mm. This relatively large error required for detection (approximately 10%) is consistent with our reaching results, which show that IPD errors have a smaller than expected impact on reach distance. At intermediate (1.3 m) and far (2.5 m) distances, psychometric functions were predominantly flat, indicating that most participants could not distinguish the error from the reference. Additionally, at the VID, the geometric model predicts zero geometric distortion for viewing errors; consequently, an objective ground truth for "correct" interval identification—which is required to extract detection thresholds—is theoretically undefined.

To confirm whether participants were performing above chance, we applied a one-tailed binomial test to the psychometric fits. The results confirmed widespread insensitivity to IPD mismatches. Out of 64 fits at the far and intermediate distances, only 7 exhibited performance significantly above chance at the far distance resulting in an IPD detection threshold of >41 mm. Insensitivity persisted for some even at 0.5 m, where stereo cues are strongest; 13 participants failed to exceed chance performance in at least the positive or negative IPD error conditions.

In contrast, participants were consistently sensitive to direct passthrough errors (Figure 6A). Median thresholds were between 1.8–3.0 mm across viewing distances for positive errors, and 6.1-8.3 for negative errors. Ulike IPD errors, the failure rate was minimal (only 16 failures total across all conditions), indicating a much more robust perceptual signal.

*Scene Complexity.* In the naturalistic forest scene, depth cues other than stereopsis provide reliable distance information. In our sparse scene, where these other cues are removed, we observed that sensitivity to IPD errors improved dramatically, whereas sensitivity to direct passthrough errors decreased. Most notably, IPD viewing errors at the far distance were largely undetectable in the natural scene (55 of 64 psychometric functions), but detection rates improved significantly in the sparse scene (dropping to 28 of 64 undetectable cases). For participants who met the inclusion criteria in the sparse condition, median IPD error thresholds were 5.0 mm (negative) and 6.8 mm (positive) - a massive improvement from the 41.4 (negative) and 100 (negative) mm thresholds obtained from the 9 thresholds that met the inclusion criteria in the complex scene. At the near distance median IPD thresholds improved from 5.6 mm and 5.4 mm (natural) to 4.6 mm and 4.9 mm (sparse) for negative and positive errors, respectively. These results indicate that perceptual sensitivity to IPD errors is strongly dependent on scene content and increases when other depth cues are eliminated.

Conversely, sensitivity to direct passthrough rendering errors followed the opposite trend: median thresholds were substantially elevated (indicating lower sensitivity) across most conditions in the sparse scene. This is expected because the geometric impact of passthrough error diminishes as viewing distance increases. In the naturalistic scene, there are still objects located close to the viewer remain even when the target object is farther away, providing high-sensitivity references. The removal of these foreground objects in the sparse scene resulted in an approximately 3× increase in thresholds at the far distance compared to the near distance.

## 6 DISCUSSION

*Headset Optics.* Our study purposefully ignores the impacts of near-eye optics and treats the display stack as an idealized, perfectly head-locked planar display at a fixed VID. This is necessary abstraction in order to generalize our findings to all HMDs, but ignores the realities of real HMD optics. Our simplification treats viewing errors as purely as geometric translations of the view frustum, and this approach isolates the impact of geometric perspective errors while purposefully excluding optical effects. Real viewing errors with near-eye optics are a more complex interaction between the actual displacement from the nominal eye position and the optical design which introduces pupil swim and other optical aberrations. It is important to account for these differences when considering the impacts of viewing errors in real devices.

*Implications for IPD Adjustment.* Our results indicate that reach behavior remains robust even in the presence of significant IPD errors (±12 mm). We found that IPD offsets are generally imperceptible except at large magnitudes and close viewing distances (6 mm at 50 mm viewing distance). This aligns with prior research demonstrating a general insensitivity to absolute IPD mismatches in stereoscopic content [Allison and Wilcox 2015; Didyk et al. 2011; Guan and Banks 2016; Stevenson et al. 1989]. While IPD shifts are known to alter the perceived scale of objects [Kim and Interrante 2017; Mine et al. 2020; Piumsomboon et al. 2018], their specific contribution to distance judgment remains unclear. However, perceptual tolerance does not eliminate the need for mechanical IPD adjustment. Physical IPD and IAD alignment is still required to mitigate view-dependent optical artifacts (such as blur and distortion), even if the user is geometrically insensitive to the perspective shift.

*Implications for Video Passthrough.* While current commercial HMDs typically rely on static rendering pipelines, our results suggest that the geometric viewing errors inherent to these systems can induce perceptual inaccuracies. This is particularly critical for video passthrough HMDs. Unlike standard VR, where the virtual camera can be placed anywhere, passthrough cameras are physically constrained by the displacement between the external cameras and the user's eyes. Our findings support the adoption of computational imaging methods [Kuo et al. 2023] or view-synthesis algorithms to correct this parallax mismatch [El Chemaly et al. 2025; Xiao et al. 2022]. However, implementation requires a careful trade-off: the latency and potential image artifacts introduced by these methods can sometimes generate registration errors more severe than the original static offset. Consequently, alternative hardware approaches—such as reducing the physical camera-to-eye distance via thinner optics [Maimone and Wang 2020]—may act as a necessary complement to software correction to ensure perceptual accuracy.

*Active vs. Static Observers.* Our study focuses on static observation to isolate specific geometric variables; however, real-world HMD usage involves continuous head and eye movement. Perspective projection errors that are imperceptible to a static observer often manifest as dynamic distortions—such as world instability or "swimming"—when the user moves [Guan et al. 2023]. These artifacts arise from discrepancies in motion parallax, a depth cue that can supersede stereoscopic cues in sensitivity. Consequently, the detection thresholds reported here may decrease with certain head and eye movements. Future work should extend this protocol to active observers to quantify the true limit at which geometric errors disrupt the stability of the virtual environment.

## 7 CONCLUSION

We highlighted the perceptual consequences of rendering and viewing errors in HMDs To do this we first introduced a framework that differentiates between errors in stereo geometry introduced during rendering and viewing stages which can be clearly visualized in an accompanying interactive web application. We next built a software platform that corrects for naturally occurring and unavoidable viewing errors in a Quest 3 HMD which can then be used to purposefully simulate rendering and viewing errors. This platform enabled the set of blind reaching and two-interval forced choice experiments investigating the perceptual consequences of three ecologically-valid rendering and viewing errors in HMDs - a set of experiments that had not been previously possible despite their direct relevance to all egocentric, head-tracked display architectures.

We showed that errors in stereo geometry can induce changes in blind reach behavior but these blind reach distances need to be transformed to correct for reaching bias and put in the egocentric coordinate frame to be interpreted as perceived distance. Distance misperceptions induced by direct passthrough and eye relief errors are well predicted by our geometric model, but depth error induced by IPD error is smaller than predicted by geometry. Evaluating the detectability of IPD error in a 2IFC task revealed that, for a naturalistic and cue-rich scene, even detection thresholds for IPD errors was relatively high at 50 cm, and imperceptible to most viewers at 2.5 m. Overall, our findings demonstrate the perceptual impact of rendering and viewing errors in VR.

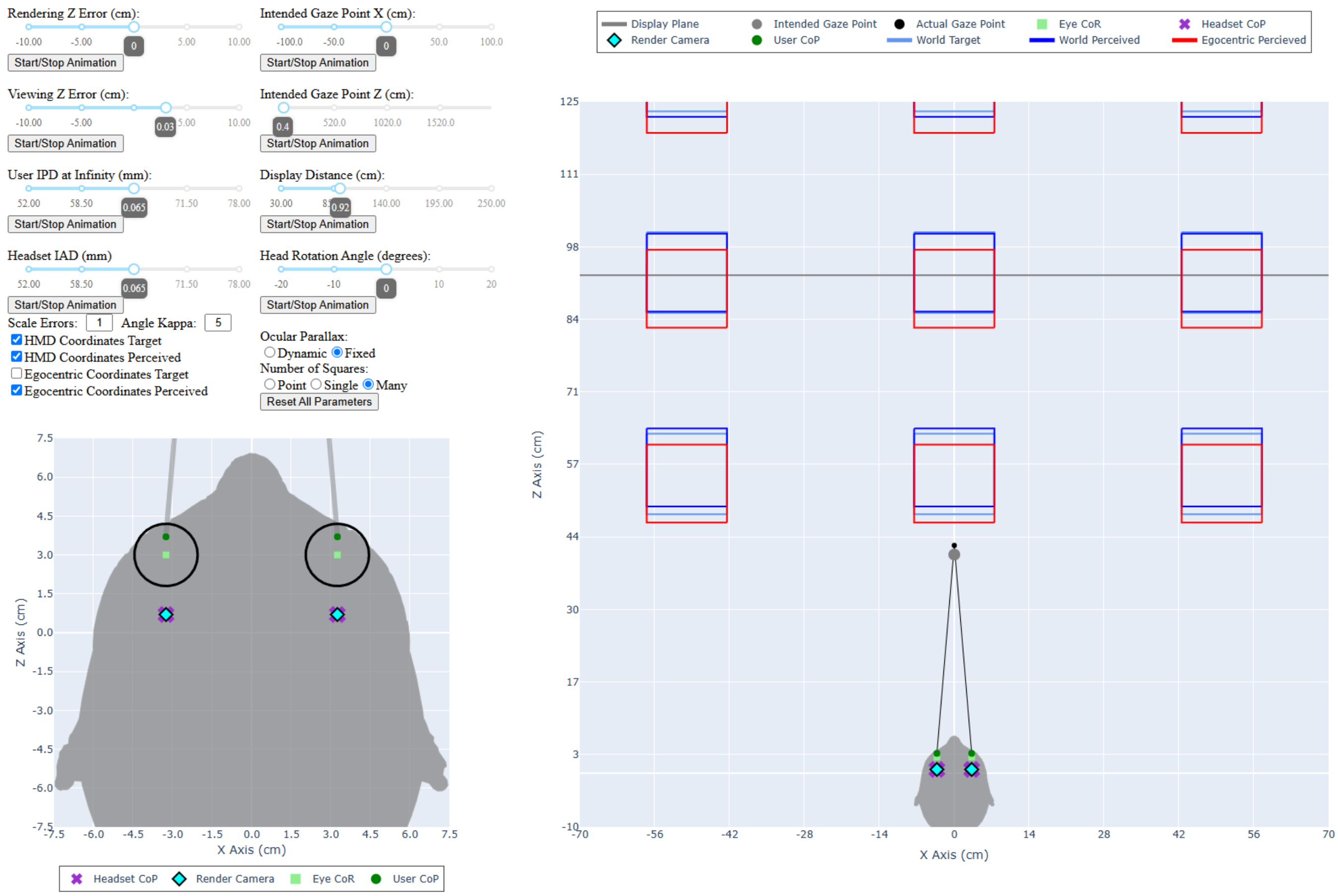

Fig. 8. Interactive perspective stereo geometry visualizer provided in supplementary materials. In this simulation a viewing error moves the viewer 3 cm forward relative to the HMD coordinate system which leads to a 3 cm offset between the egocentric and HMD coordinate systems. This leads to potential discrepancies in conclusions about perceived distance if measurements taken in these two coordinate systems are directly compared. The intended render geometry in the HMD coordinate system is shown in light blue. The perceived geometry in HMD coordinate system according to perspective geometry is shown in dark blue. For viewing errors, objects on the virtual image plane are always perceived to be on the plane, according to stereo geometry. This means the center row of objects will appear near their intended position. However, because the viewer is closer to the display by 3 cm, this means that they will perceive the objects to be closer to them by 3 cm (boundaries in red) compared to the assumed HMD render geometry. This difference in coordinate systems can lead to seemingly paradoxical situations. For the row of objects closet to the user, the user will blind reach to the dark blue boundary, leading an external observer to think distance is overestimated (compared to light blue target). Yet the actual perception of the user is the red boundary and distance is in fact underestimated.

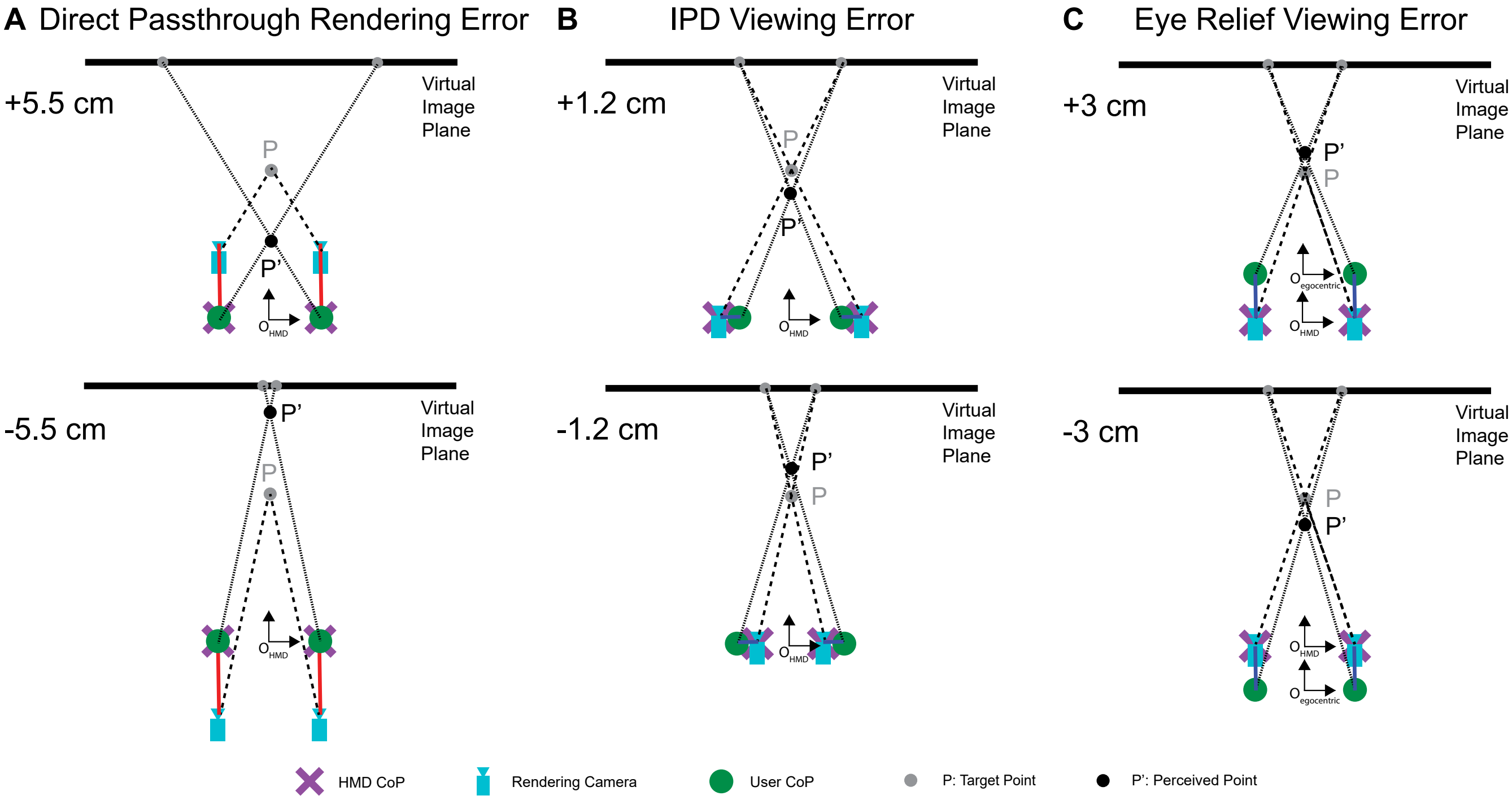

Fig. 9. Rendering and viewing error conditions for the blind reaching experiments. **(A)** Direct passthrough rendering error results from showing images of cameras attached to the front of HMD without view reprojection. **(B)** IPD viewing error occurs when interaxial distance (IAD) of the HMD is smaller or larger than the user IPD. **(C)**. Eye relief viewing error is due to user's eyes being in front or behind the HMD CoP due to facial geometry.

# Perceptual Sensitivity to Stereo Geometry Errors in Head-Mounted Displays

Supplementary Material

RAFFLES XINGQI ZHU, Reality Labs Research, Meta, USA McGill University, Canada
CHARLIE S. BURLINGHAM, Reality Labs Research, Meta, USA
OLIVIER MERCIER, Reality Labs Research, Meta, USA
PHILLIP GUAN, Reality Labs Research, Meta, USA

## 1 STEREO GEOMETRY MODEL

### 1.1 Additional Details on Stereo Triangulation

*Rendering Errors.* The first step of an HMD render pipeline is to capture images, either from render cameras placed in a virtual scene, or from physical passthrough cameras on the HMD. We define rendering error as displacement of the render camera (or video passthrough camera) relative to the HMD CoP. For rendered content these errors generally arise from headtracking inaccuracy and user movement that moves the HMD CoP from the headset pose used for rendering (i.e., latency). Its impact is illustrated in Figure 1B. Here the HMD CoP is on-axis with a sphere, but the render camera is not. Rays are captured by the render camera at an angle and the image presented on the display is an oval. The user views the display from the HMD CoP and perceives an oval rather than a circle which is the retinal image what a user would have seen if they had viewed a real sphere.

In video passthrough systems without view reprojection [El Chemaly et al. 2025; Kuo et al. 2023] a static offset will exist between the cameras located at the front surface of the headset and the headset CoP. This type of passthrough error is similar to enlarging the user's head which will effectively exaggerate rotational head movement, corresponding to users viewing the world as if their heads were larger. Importantly, rendering errors affect parallax when user's move their heads, and objects closer to the rendering camera in a virtual scene are more affected by rendering errors compared to farther objects.

*Viewing Errors.* Rendered or captured content is presented on the actual HMD displays from the left and right HMD CoP. These displays are very close to the user's eyes, and near-eye optics magnify these displays and move them to a comfortable virtual image distance, a process which requires optical distortion correction [Robinett and Rolland 1992; Rolland and Hopkins 1993]. For simplification and generalizability we assume that the near-eye optics in HMDs are perfect lenses, and that images captured by the render camera can be simply projected onto a wide FOV plane at a virtual image distance (VID), typically between 1-2 meters away from the HMD CoP.

If the CoP of the user's eyes are not at the HMD CoPs then the images on the displays will not be perceived as intended. We define viewing error as displacement of the user eye CoP relative to the HMD CoP. An example of viewing error can occur if a user wears their own corrective glasses in an HMD which increases eye relief and place the user's eye too far away from the HMD CoP. In Figure 1C the render camera is correctly located at the HMD CoP so the perspective image of the sphere is a circle. However, the user's eye is not at the HMD CoP so they see an oval instead. In this scenario the user's eye is off-axis to the sphere so they would see an oval in real life as well. However, the parallax shifts seen by the user during a viewing error is dependent on the virtual image distance, rather than the object distance. This underscores one of the important distinctions between rendering and viewing errors — objects very far away are marginally affected by rendering errors, but because the VID of an HMD is typically between 1-2 meters viewing errors can have large impacts on the perceived distance of these far objects.

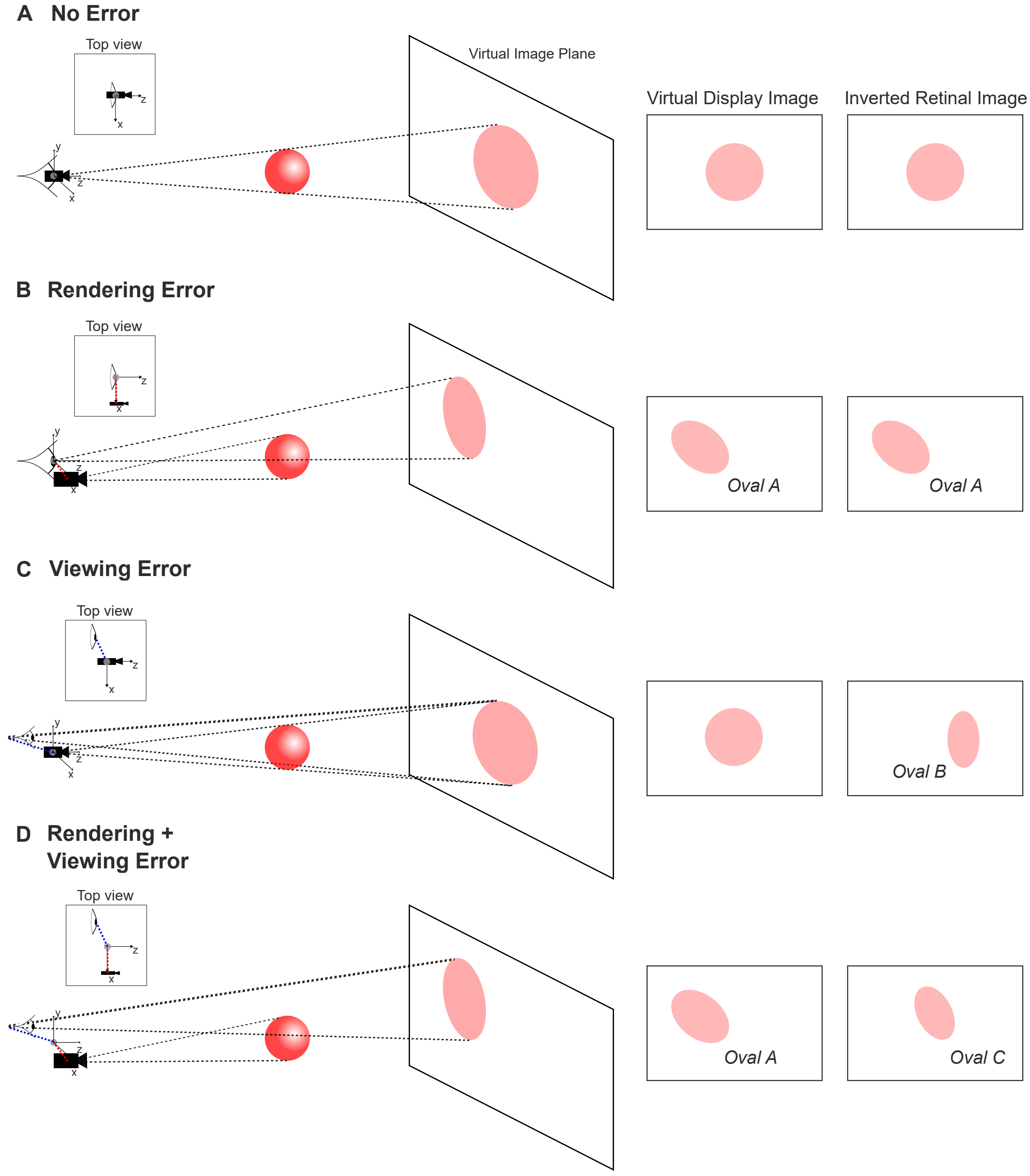

Fig. 1. Comparing rendering error with viewing error (copied from main text). For simplicity, we show a monocular example as the set up is the same for both eyes. The origin of the coordinate system is at the HMD CoP (grey dot). The virtual sphere is positioned directly in front of the the HMD CoP. **(A)** In the ideal scenario, the render camera, CoP of the user's eye, and HMD CoP are all at the same position in space. No rendering or viewing error exists in this system. The camera is on-axis with the sphere and will render the corret perspective projection (a circle) that will be presented on virtual image plane. The user will view this circle from the intended CoP and the user's retinal image of what is shown on the display plane is also a circle. **(B)** The render camera is displaced to the right and down from the HMD CoP, creating a rendering error. The camera will render the sphere from this incorrect location, and render an oval instead of a circle. This oval is shown on the display plane and the user views it from the correct position so they perceive the same incorrectly rendered oval. **(C)** The CoP of the user's eye is displaced to the left and behind the HMD CoP, creating a viewing error. Despite the rendered image being correct on the virtual image plane, the perceived image is distorted. **(D)** In realistic usage, rendering and viewing errors typically occur together and can result in geometric errors being introduced in both render and viewing stages.

*Combined Rendering and Viewing Errors.* Figure 1D shows the combined effect of these two types of viewing errors. Here a distortion from the target perspective projection of a circle is first introduced by the render camera being displaced from the HMD CoP. The oval that is shown on the display is additionally distorted from the user eye CoP being displaced from the HMD CoP, leading to an image of an oval on the retina that is different from the oval that was initially rendered.

## 1.2 Coordinate Frame Transform of Reaching Data

Our participant reach data is recorded in the HMD coordinate frame and, for blind reaching, these distances must be transformed into egocentric coordinates to account for offsets between where the actual viewer's eyes are and where they are positioned in the HMD simulation platform. Table 1 shows the average displacement across all trials and subjects.

*1.2.1 Egocentric Coordinates.* We use the largest eye relief setting in the Quest 3 to accommodate the eyetracker added to the HMD. This results in an average eye relief that is 13 mm larger than expected (Table 1, Z coordinate) and must also be accounted to interpret our measured reach as a user-perceived reach distance (Figure 2). This is accomplished by adding each individuals' eye relief error to the controller reach distance reported by the HMD. This transform is applied to all blind reaching conditions and this coordinate frame best represents how far away the target reach object appeared to the user.

*1.2.2 Egocentric to HMD Coordinate Conversion.* In a typical HMD, egocentric reach distances cannot be measured without knowing both the actual user eye position and the nominal assumed user eye position. Thus, a more reliable measure of reaching performance is to instead consider reach distance in the HMD coordinate frame. In our HMD simulation platform, the IPD and passthrough error conditions do not displace the viewer and HMD origin. Therefore no additional transforms are necessary beyond compensation for the real eye relief errors described in Section 1.2.1 to interpret the reaching results in HMD coordinates for IPD-IAD and passthrough errors. Eye relief viewing errors require an additional transform to account for the fact that the viewer's eye is not actually at the simulated position (Figure 2) and in our study this means an additional ±3 cm shift to the egocentric blind reach distance. This coordinate system best represents the distance a user would reach to if the rendering and viewing errors simulated in our HMD platform were actually present in a real HMD instead of being simulated in our platform.

| | X (mm) | Y (mm) | Z (mm) |
|---|---|---|---|
| Left Eye | 1 ± 2 | 0 ± 3 | -13 ± 3 |
| Right Eye | 0 ± 2 | 0 ± 3 | -13 ± 3 |

Table 1. Average entrance pupil position relative to the HMD COP of Quest 3. n = 6295 trials.

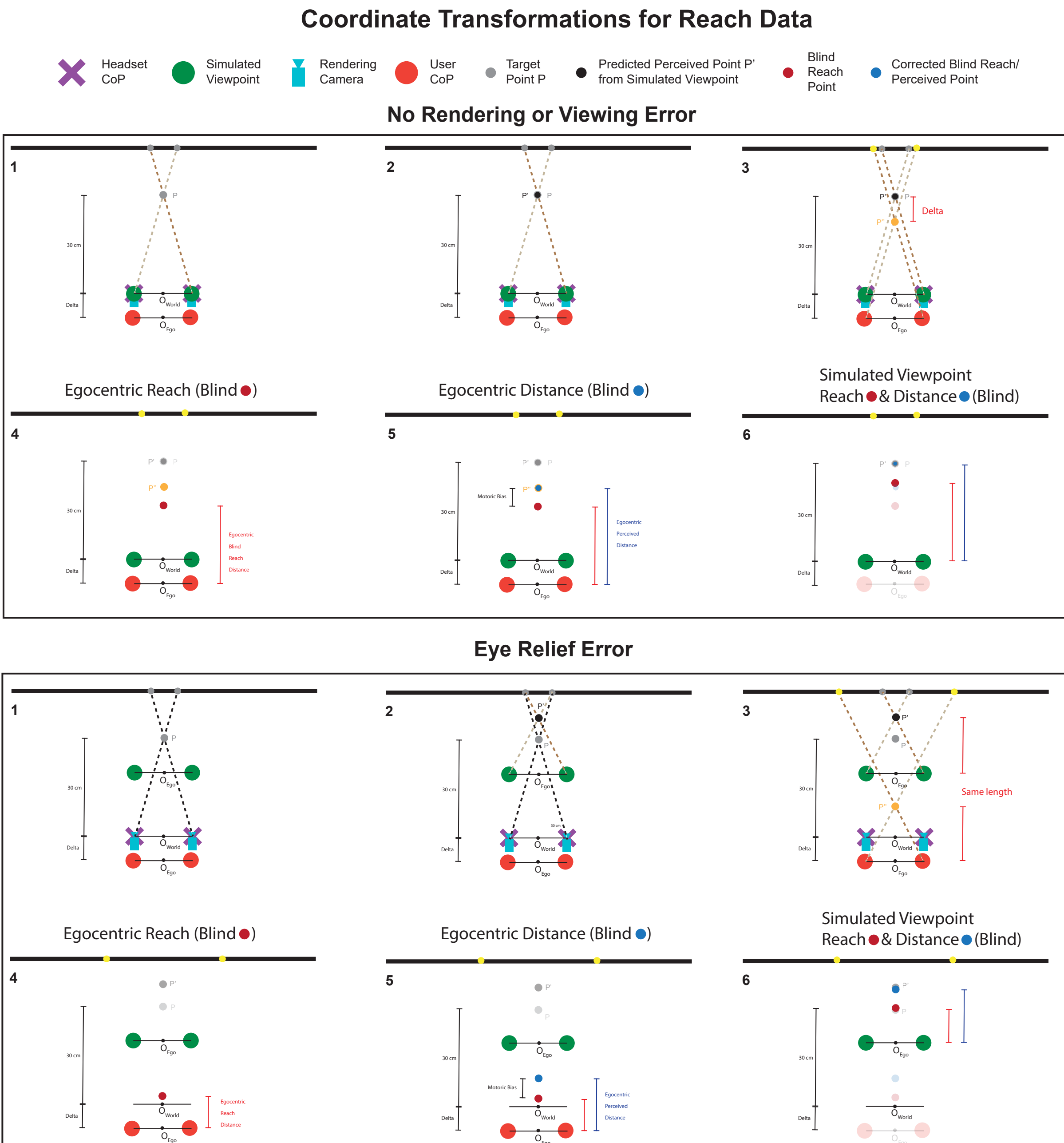

Fig. 2. Coordinate system transforms used to interpret reaching data studies. (Top) Our headset uses the largest Quest 3 eye relief setting available which results in an average eye relief viewing error of -12 mm. In panels 1-2, we show how a target point P is rendered at P' when no stereo geometry errors are added to the system. In panel 3, the rays used to simulated viewpoint are shown to the actual user who is slightly displaced away from the ideal eye relief resulting in a perceived point P'. In panels 4-6, we describe how blind reaching and bias-corrected blind reaching can be converted from measured values into their equivalent perceived values based on the small offset delta between the actual user eye position and simulated viewpoint. (Bottom) Simulated eye relief viewing errors require another transform to interpret egocentric reach data in the simulated viewpoint world coordinates which is shown in panel 5 and 6.

## 1.3 Geometric Distortion Fields

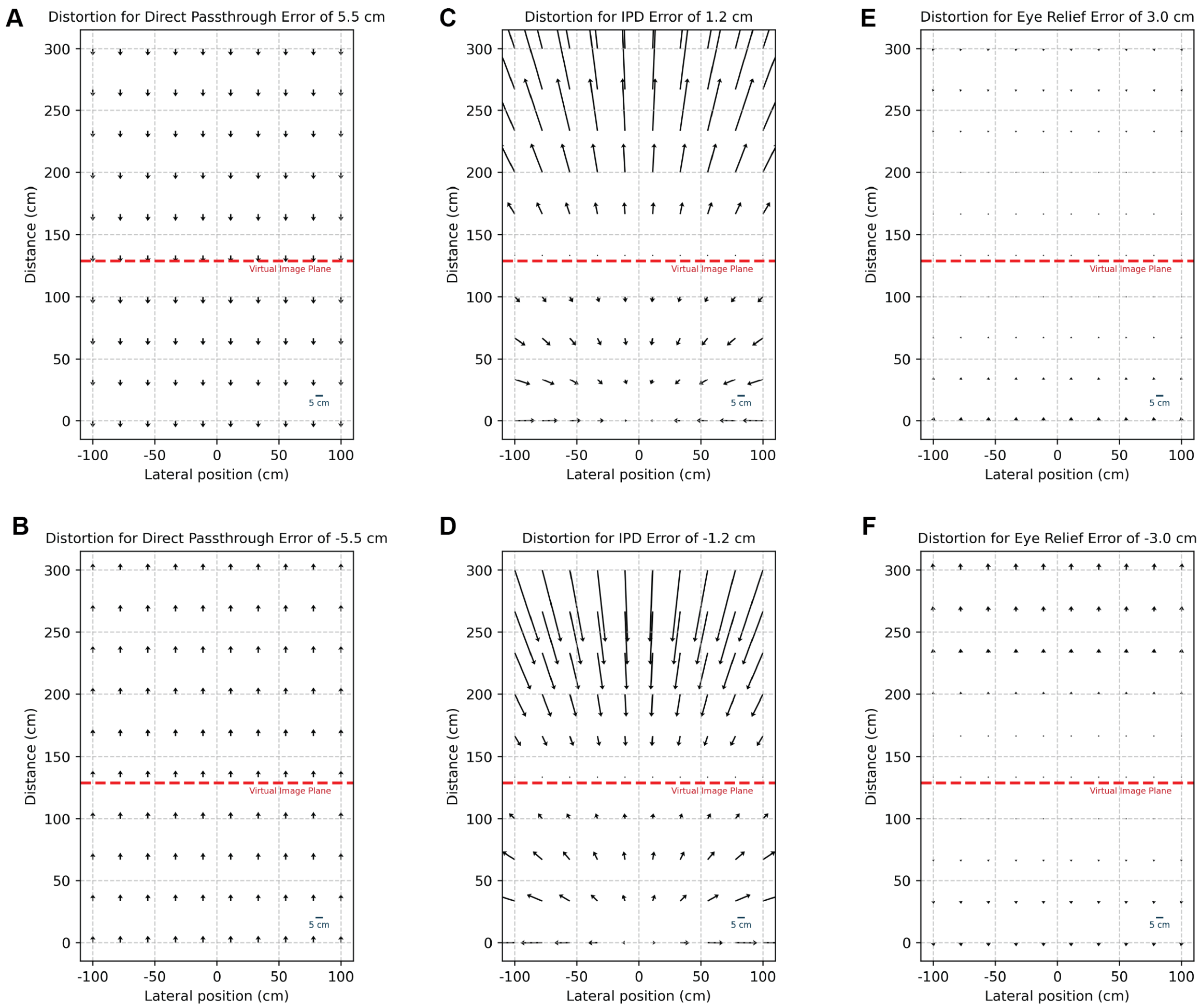

Fig. 3. Geometric distortion fields in HMD coordinates at errors used in reaching experiments. (A,B) Direct Passthrough errors. (C,D) IPD Error. (E,F) Eye Relief Error. For viewing errors (C-F), points on the virtual image plane are not distorted. For rendering errors (A-B), distortions are the same irrespective of virtual image plane distance. Ocular parallax is on for panels A and B to visualize the effect of rendering error only.

## 2 EXPERIMENT 1

### 2.1 Reaching Experiments Supersubject Raw Data

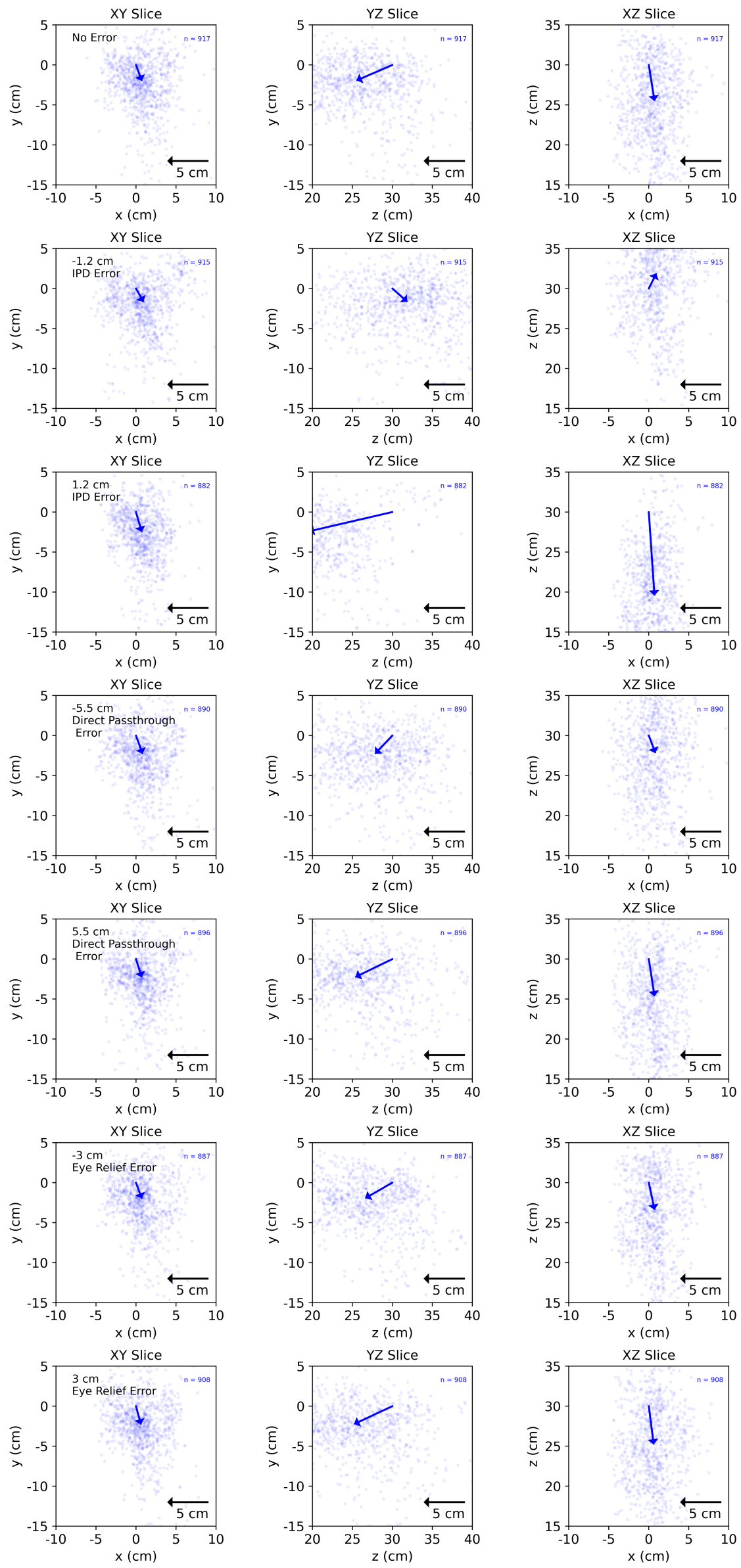

Fig. 4. Raw recorded reach positions across all 32 participants. Each row represents the same data projected onto different planes, visualizing from the front (XY slice), side (YZ slice), and top (XZ slice). Each dot is a single participant's reach. Arrowhead coordinates are determined from the medians of x, y, z reach positions separately. The coordinate system is left handed. Physically, positive x, y, z, correspond to the right, top, and front of the user, respectively.

## 2.2 Reaching Experiments One Example Subject Data

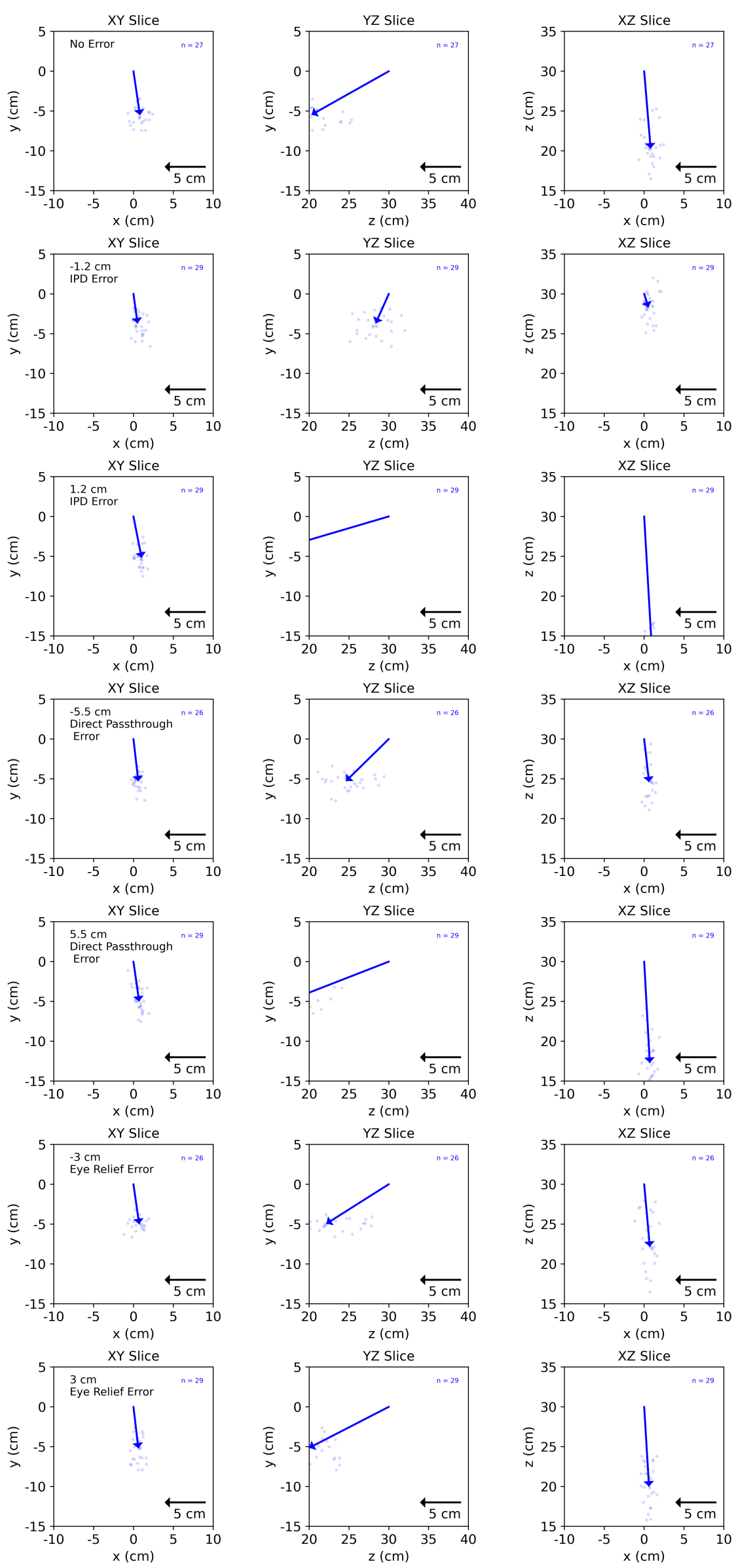

Fig. 5. Raw recorded reach positions for a single participant. The XZ Slice of No Error condition is shown in Figure 1B of the main text.

# 3 EXPERIMENT 2

## 3.1 Geometric Model Predictions

Geometric simulations of IPD viewing error predict that the perceived distance exhibits significantly different behavior based on the target object's distance relative to the VID. If the object is in front of the VID, then the perceived object distance will be farther than intended when the HMD IAD is smaller than the user IPD (Figure 6B,C). If the object is behind the VID, then the same error will lead to the perceived object distance be closer than the intended distance (Figure 6B,E). If the object is at the VID exactly, then IAD and IPD mismatches should have minimal impact on the object's perceived distance (Figure 6B,D). For direct passthrough rendering error, we do not expect this behavior. A positive error (render camera in front of HMD CoP) make the stimulus appear closer at all three distances (Figure 6A,C-E). Our analysis in Section 3.7 shows the direct of misperception in the human data are consistent with this geometric model. Crucially, we see the magnitude of distance perception is much larger for IPD error than direct passthrough error at 0.5 m and 2.5 m based on stereo geometry.

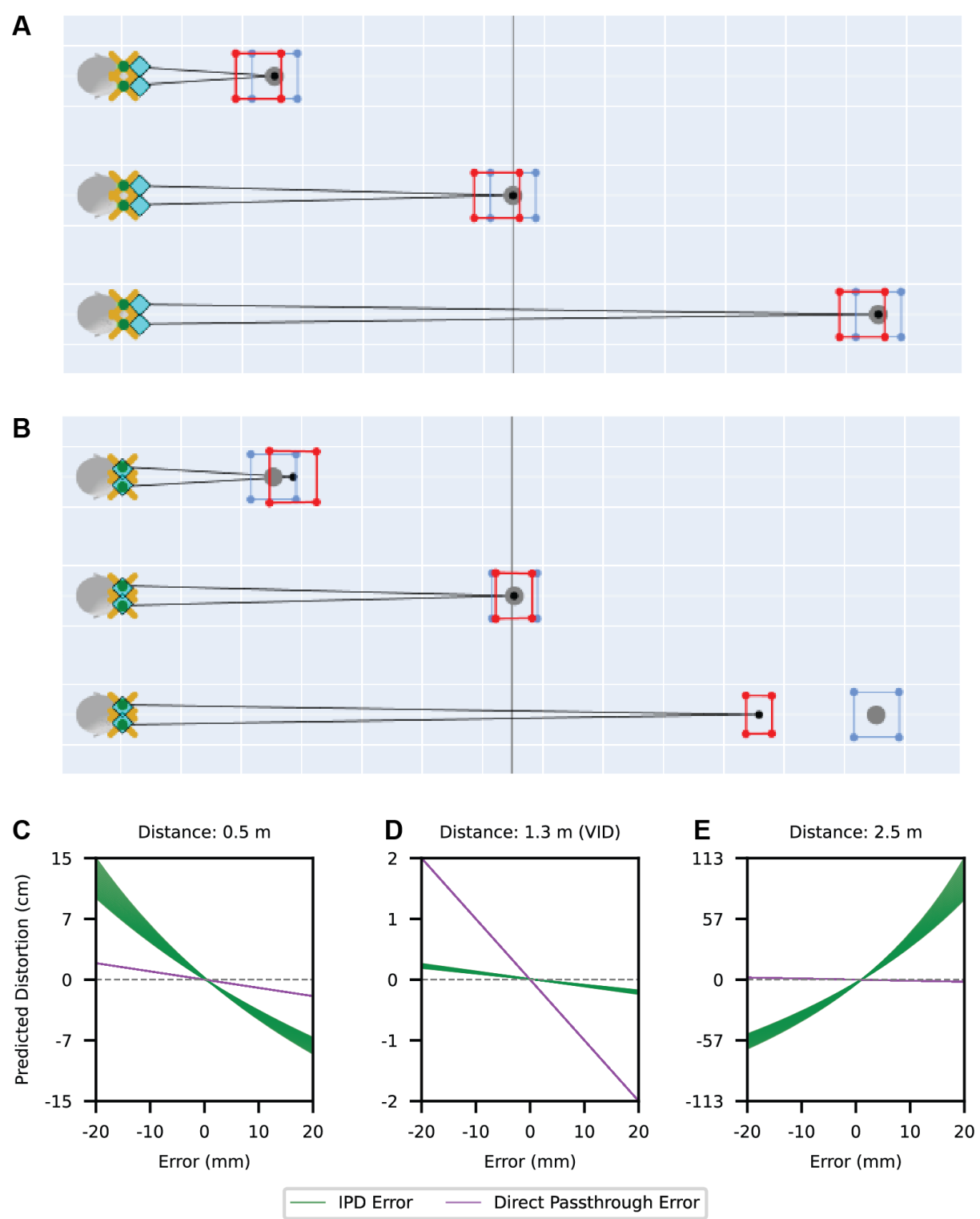

Fig. 6. Distance misperception in relation to virtual image distance (VID). **(A)** and **(B)** show triangulation model predictions for a direct passthrough rendering error of +5.5 cm (render cameras in front of HMD CoP) and IPD viewing error of -1.2 cm (HMD IAD smaller than user IPD) for target distances at 0.5, 1.3 (VID), and 2.5 meters. The target boundary is shown in blue and the perceived boundary in is shown in red. **(C)**-**(E)** Distance misperception as a function of error magnitudes for user IPDs ranging from 55 mm to 75 mm. A negative value means the object is perceived as closer.

## 3.2 Psychometric Fitting

The psychometric function we used is the logistic function:

$$P_C = \gamma + (1 - \lambda - \gamma) \cdot \frac{1}{1 + e^{-s(x-\alpha)}} \quad (1)$$

The lapse rate $\lambda$ and guess rate $\gamma$ were set to 0.01 and 0 respectively. Threshold $\alpha$ and slope $s$ were the free parameters. We used the *minimize* function of the *optimize* module from the Scipy library in Python to perform a Maximum Likelihood Estimation on the binary response data [Kingdom and Prins 2016]. We minimized the negative log likelihood of data given the model:

$$\log(L) = \sum_x [n_C(x) \cdot \log(P_C(x)) + (n_T(x) - n_C(x)) \cdot \log(1 - P_C(x))] \quad (2)$$

where $P_C$ is given by Equation 1, and $n_T$ and $n_C$ are respectively the total number of trials and the number of correct trials for a stimulus level. We bootstrapped the raw data and refit the function 200 times for each participant to obtain a distribution of the fitted threshold parameter. We found the bootstrapped threshold distributions were well behaved with no extreme values.

## 3.3 Supersubject Psychometric Function for Error Detection: IPD Error

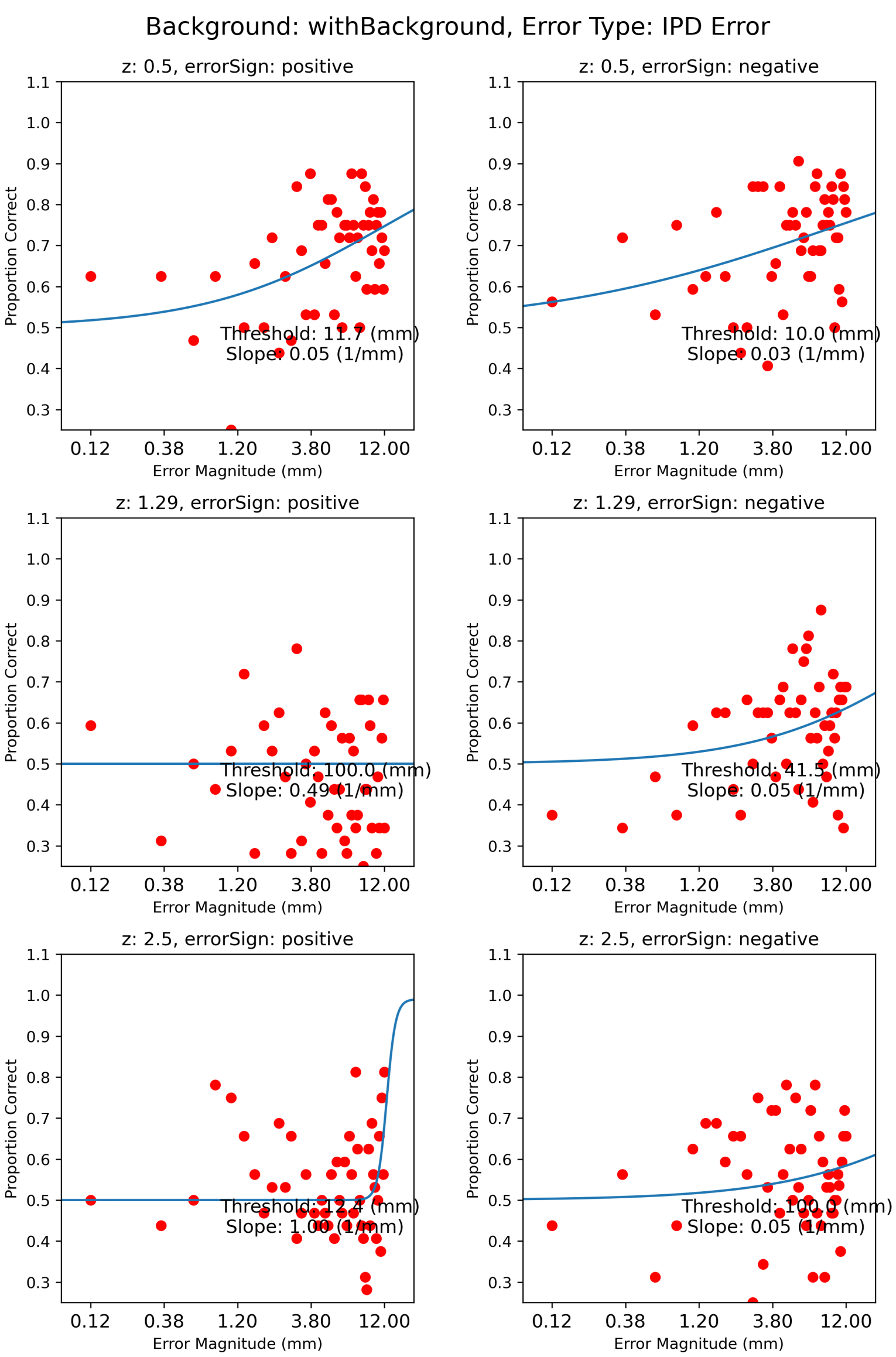

Fig. 7. Psychometric function of best fit on the combined data across all 32 participants for IPD error. Top row: 0.5 m. Middle row: 1.3 m. Bottom row: 2.5 m. Left column: positive error. Right column: negative error. Each red point represents the probability of detection, calculated from responses from 32 participants for that error level. The psychometric function is fit to the raw binary data.

## 3.4 Supersubject Psychometric Function for Error Detection: Direct Passthrough Error

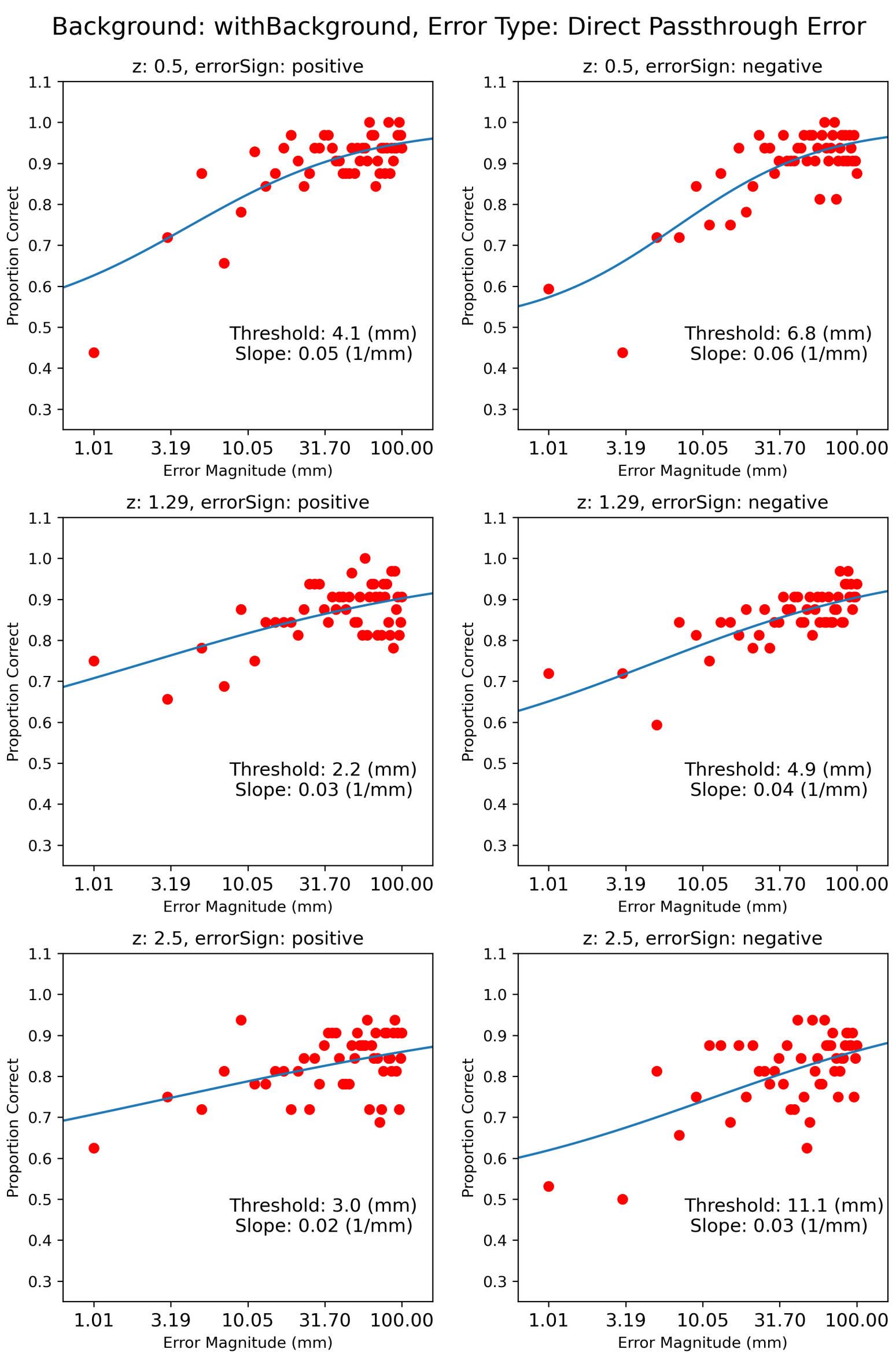

Fig. 8. Psychometric function of best fit on the combined data across all 32 participants for direct passthrough error.

## 3.5 Example Participant Psychometric Function for Error Detection: IPD Error

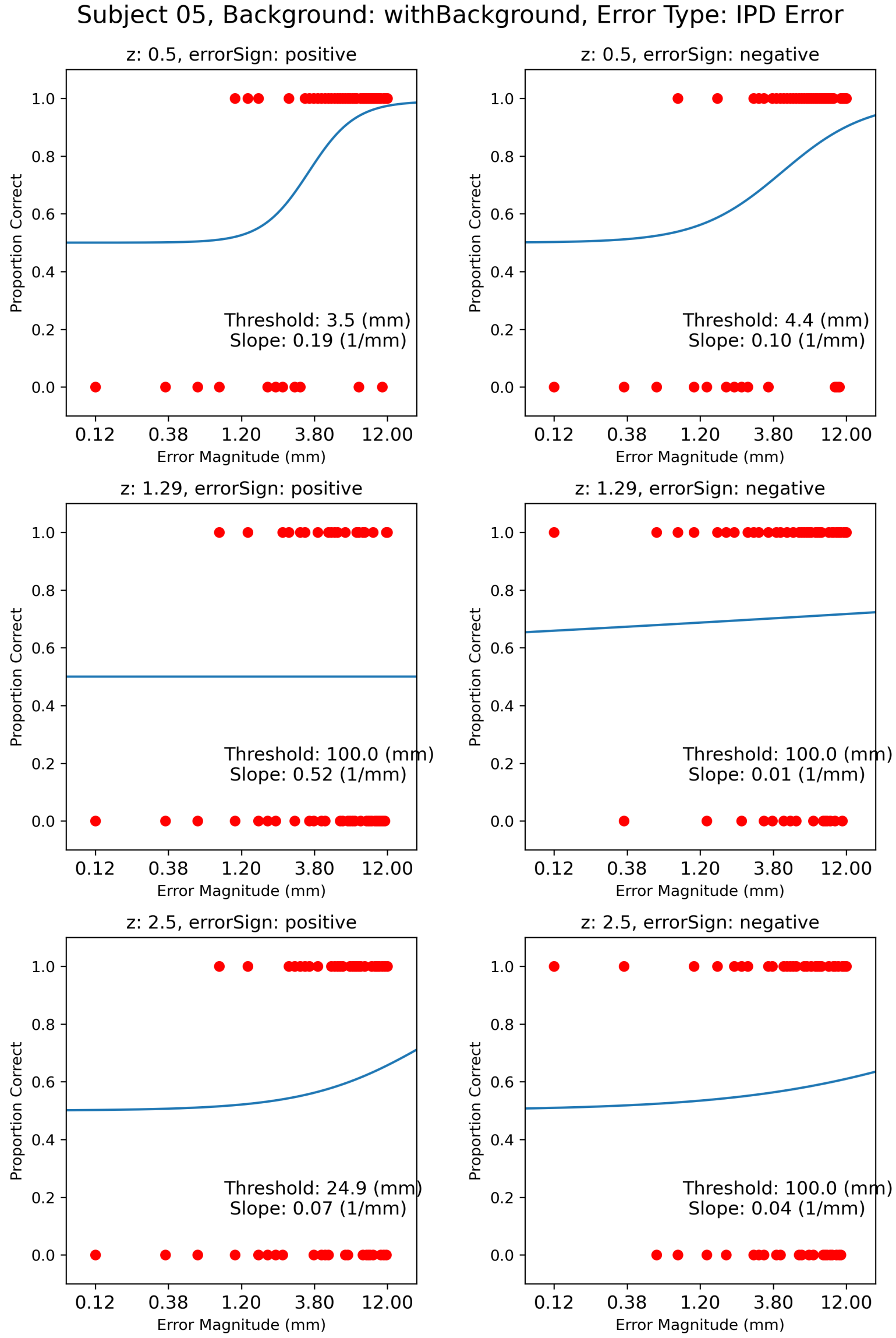

Fig. 9. Psychometric function of best fit for a single participant for IPD error. The top left panel is shown in Figure 1C of the main text.

### 3.6 Example Participant Psychometric Function for Error Detection: Direct Passthrough Error

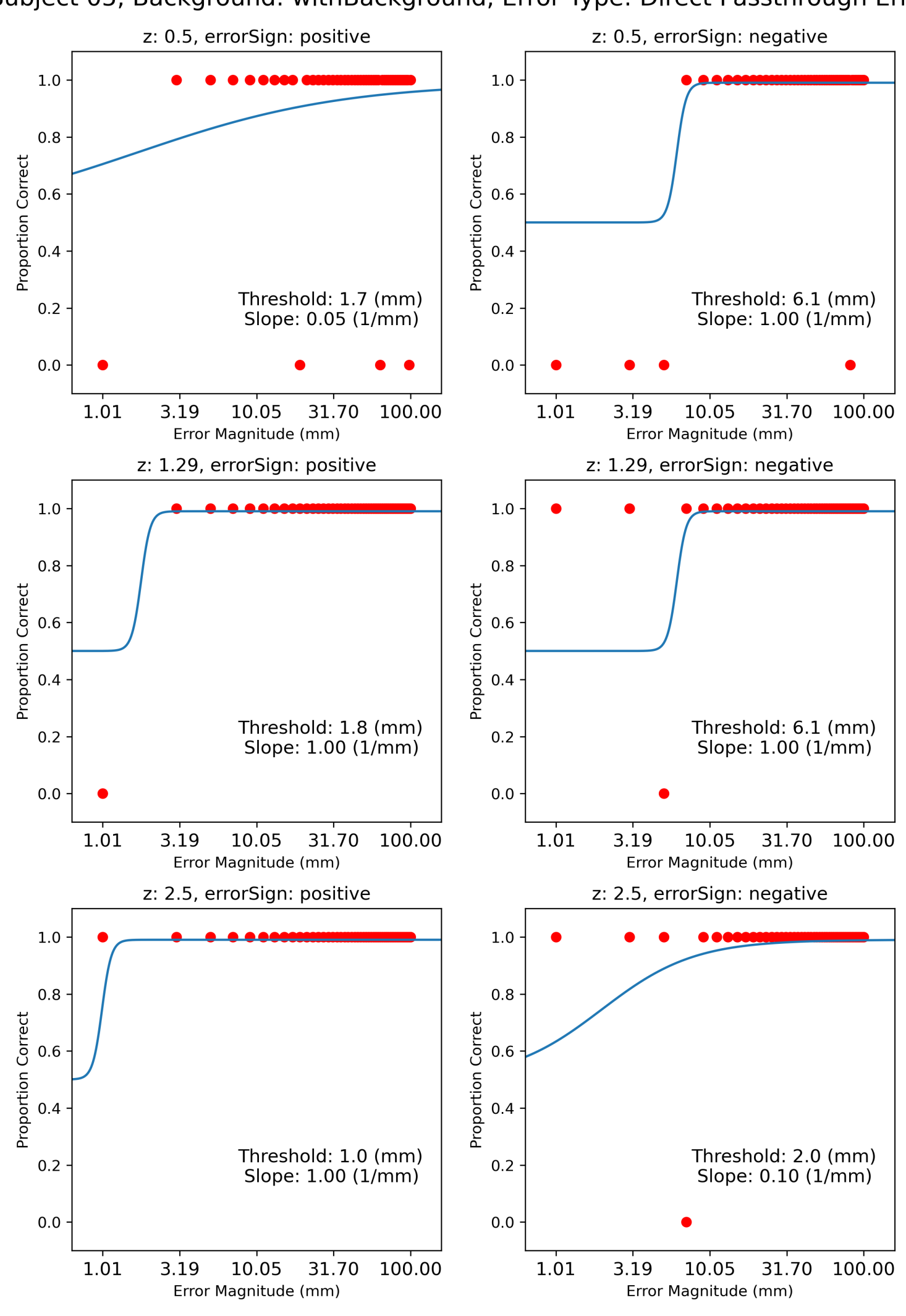

Fig. 10. Psychometric function of best fit for a single participant for direct passthrough error.

### 3.7 Psychometric Slope Analysis

To more explicitly demonstrate the change in direction of distance misperception, we carried out an additional analysis investigating the direction and sensitivity of perceived distance. This was achieved by fitting the same psychometric function described in Section 3.7 to both positive and negative error data with slope as the only free parameter with threshold $\alpha$ fixed to 0. The sign of the slope indicates how adding error affects the target's perceived distance, and its magnitude indicates the participant's sensitivity to that specific error.

We specified a maximal model linear mixed effects model to evaluate the impact of viewing distance on each participant's sensitivity to changes in passthrough rendering errors or IPD viewing errors. The maximal model in for this analysis is:

$$\text{slope} \sim \text{distance} + (\text{distance}|\text{ID}) \tag{3}$$

where slope was the best-fit slope of a single participant's psychometric function and distance was the distance of the visual target.

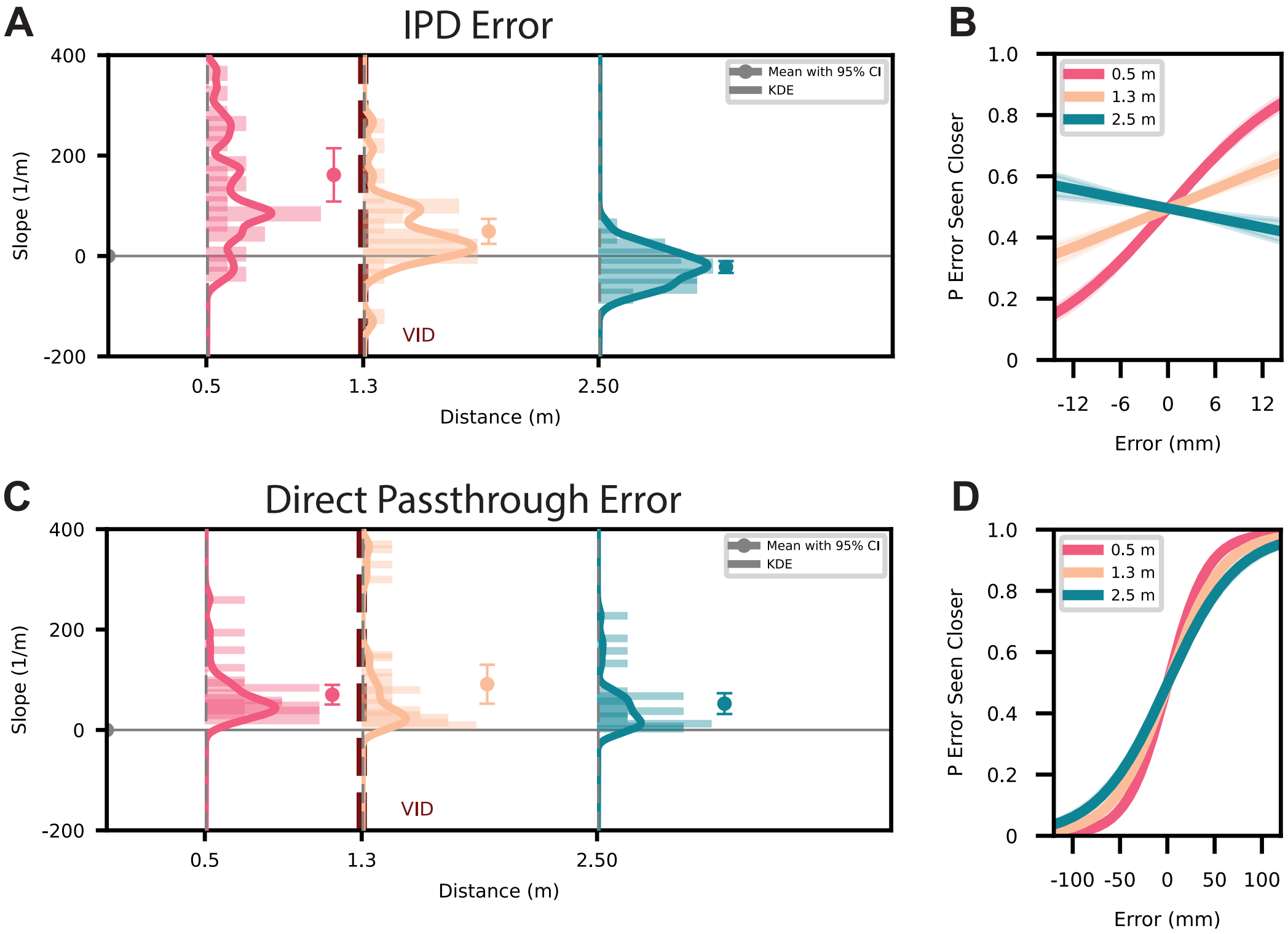

Fig. 11. Slope analysis of visual perception experiment. Panels **(A)** and **(C)** show the slope parameter values as a function of distances (0.5, 1.3, and 2.5m) for IPD error and direct passthrough error respectively. Panels **(B)** and **(D)** show the "supersubject" psychometric functions for IPD error and direct passthrough error respectively. n=32.

*3.7.1 IPD Error.* The slopes of psychometric functions fit on the combined data of all thirty-two participants (a "super subject") are shown in Figure 11B and slopes for each individual are shown in Figure 11A. At the group level, the slope did change sign for the near and far object distances (linear model estimate: -32.34 at 2.5 meters), which means that the same IPD manipulation has the opposite effect on perceived distance as predicted by the geometric model. The model predicts a zero slope when the target is at 1.3 m; we do not find a flat slope here, but we do find that the slope is flatter for the 1.3 m target compared to the 0.5 m target with more participants having psychometric functions with positive, but smaller slopes (Figure 11A). There is a clear effect where the psychometric function slope declines across the three distances (main effect of distance: $\beta$ = -88.998; t = -6.32; p < 0.0001; intercept: $\beta$ = 190.16; t = 6.20; p < 0.0001). The slopes of these functions are similar to the direct passthrough rendering errors, but appear shallower in the figures because we were limited by the range of IPD errors that could be simulated before changes in vergence demand became too significant across trials, thus the range of the IPD viewing error figures are approximately one order of magnitude narrower.

Since the psychometric function slope does become negative at 2.5 meters (linear model estimate: -32.34), it is likely that a target probe placed somewhere between 1.3 to 2.5 meters would result in a nearly flat psychometric function.One possible reason for this observed discrepancy is that objects in the scene away from the VID may influence the perceived distance to the target, more specifically, objects in front of the VID (where disparity is a stronger depth cue) could induce a general bias for objects to appear closer.

*3.7.2 Direct Passthrough Error.* Figure 11D shows supersubject psychometric functions and Figure 11C shows psychometric slopes for each participant. For all three target distances, negative errors reduced the probability that the comparison interval is seen closer than the reference interval, whereas positive errors increased the probability that the comparison interval is perceived as closer. This is more compactly represented by the slope of each psychometric function, which is positive for all three target distances (intercept in LMEM = 161.75; t = 2.42; p = 0.017; main effect of distance on slope non-significant).

## 4 CONTROLLER TRACKING VALIDATION

We placed a measuring tape on a table and compared the reported controller distance to the measuring tape and found good distance accuracy within 50 cm.

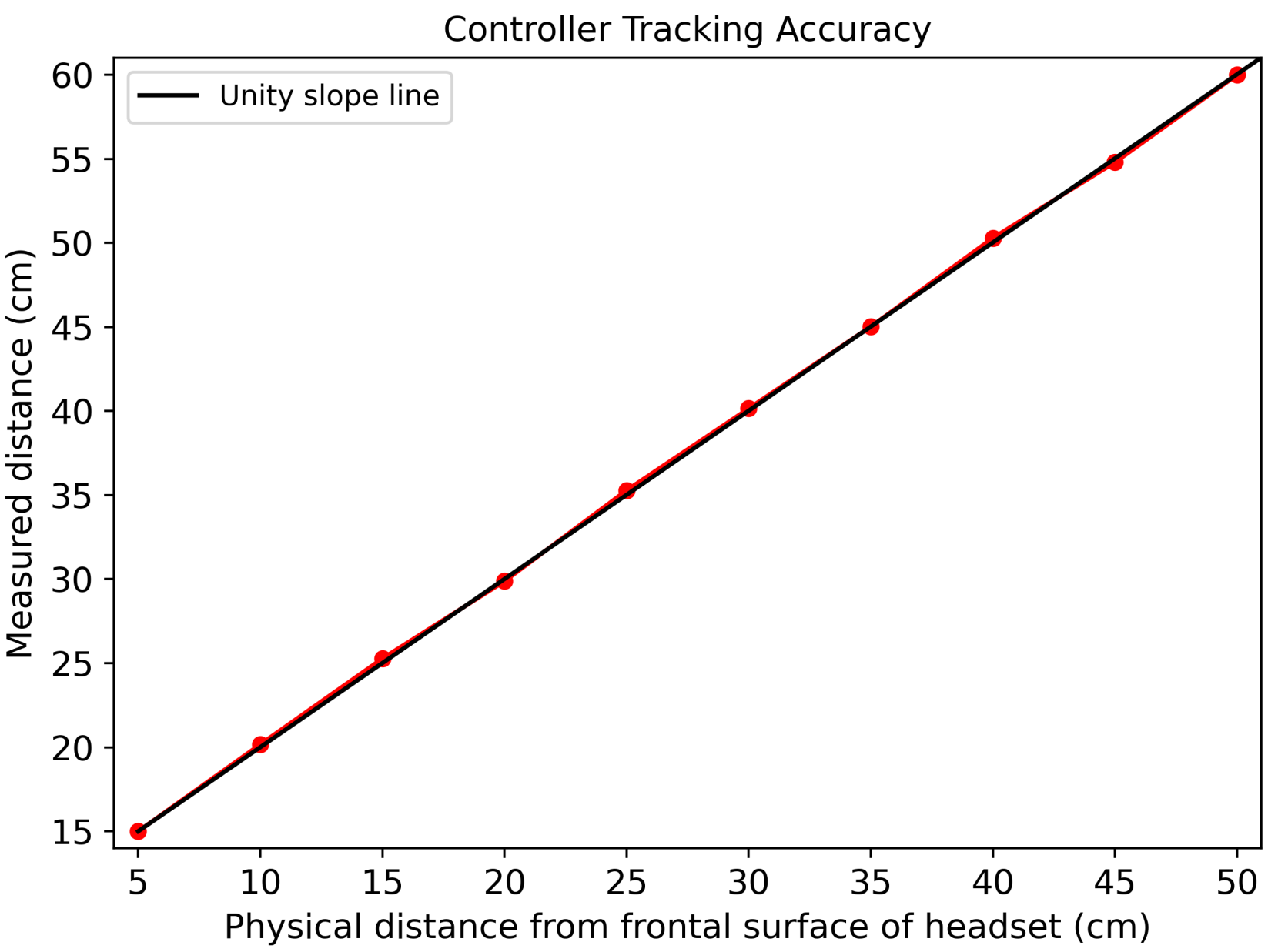

Fig. 12. Relationship between tracked controller distance and physical distance from the frontal surface of the Quest 3.

## 5 ALGORITHM FOR SIMULATING RENDERING AND VIEWING ERRORS IN UNITY

Step 1:

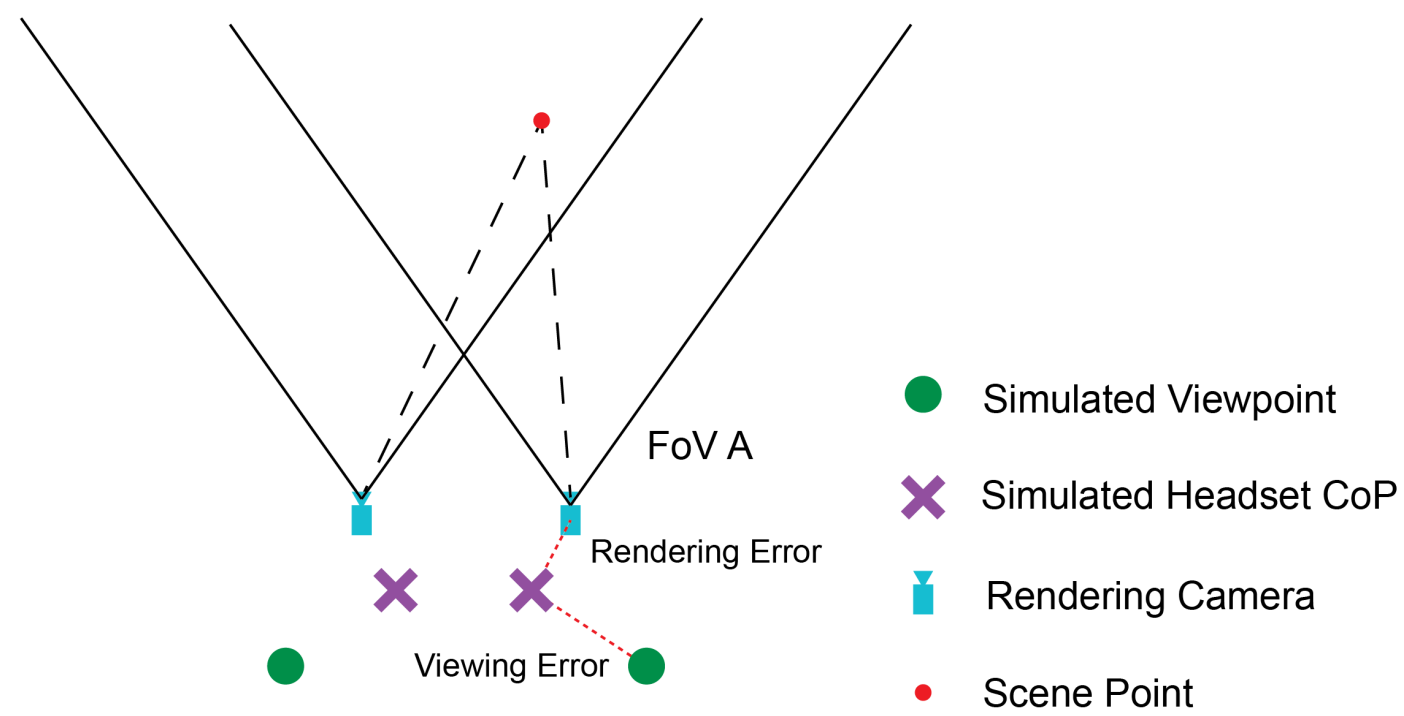

Fig. 13. Rendering camera renders the scene with field of view (FoV) A. Rendering error is the displacement of rendering camera from headset Center of Projection (CoP).

Step 2:

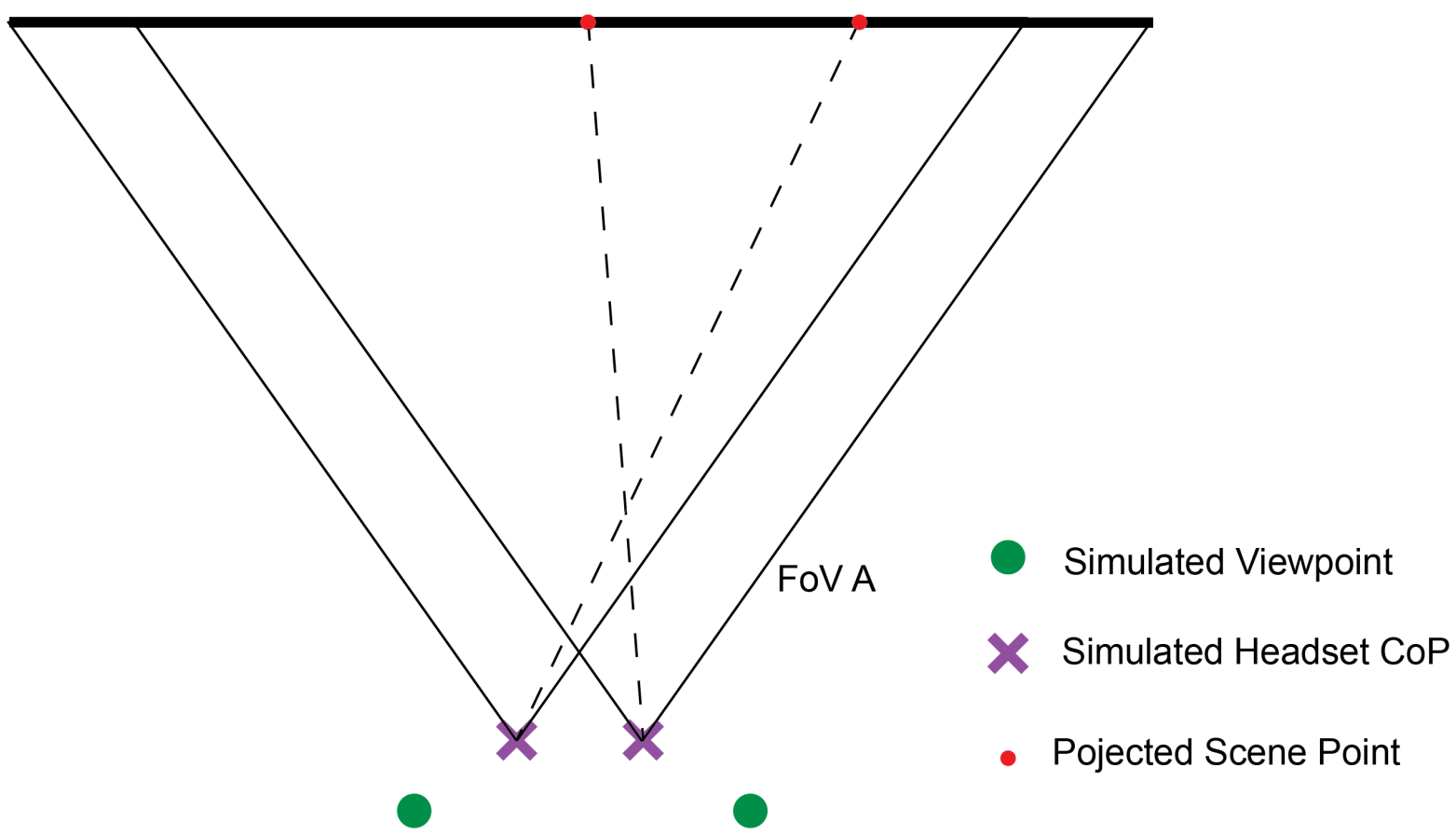

Fig. 14. Frame buffer of rendering camera is projected to a quad at the same plane as the virtual image plane. The quad size matches FoV A.

Step 3:

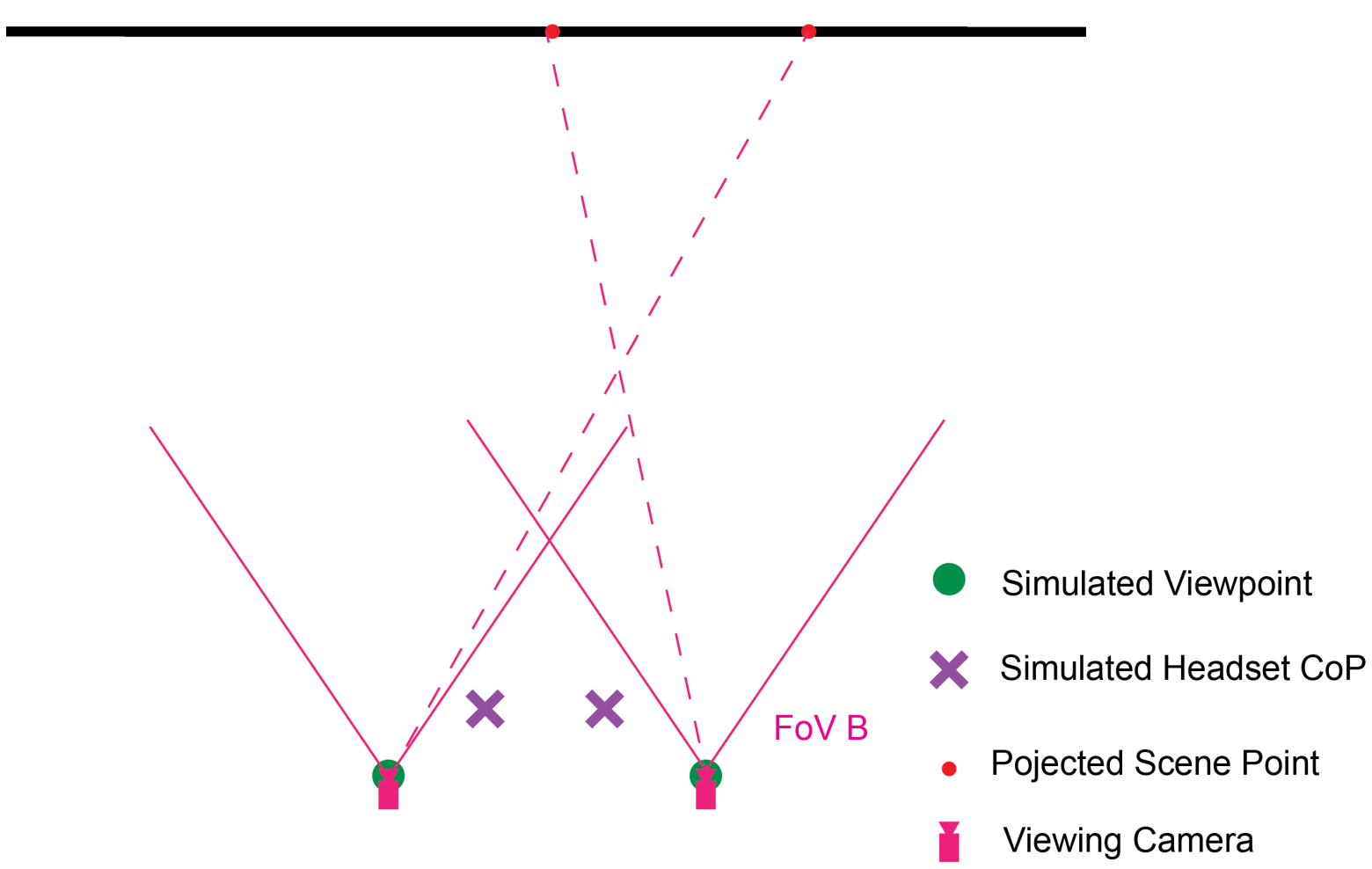

Fig. 15. Viewing camera with FoV B views this quad from the simulated viewpoint. Viewing error is the displacement of simulated user CoP from headset CoP. At parallel gaze, simulated user CoP coincides with simulated viewpoint.

Step 4:

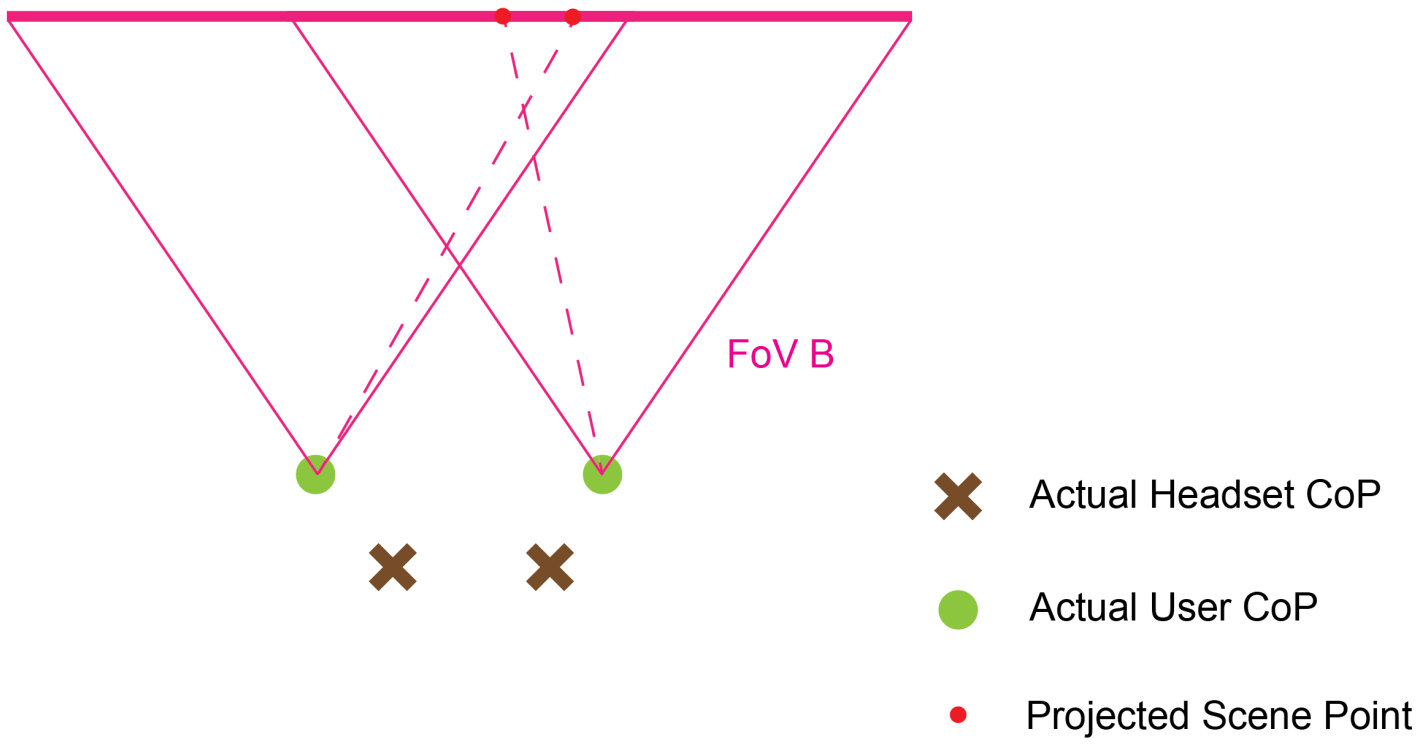

Fig. 16. Frame buffer of viewing camera is projected from the actual user CoP position to a quad. The quad size matches FoV B. We used tracked entrance pupil position to approximate actual user CoP position.

Step 5:

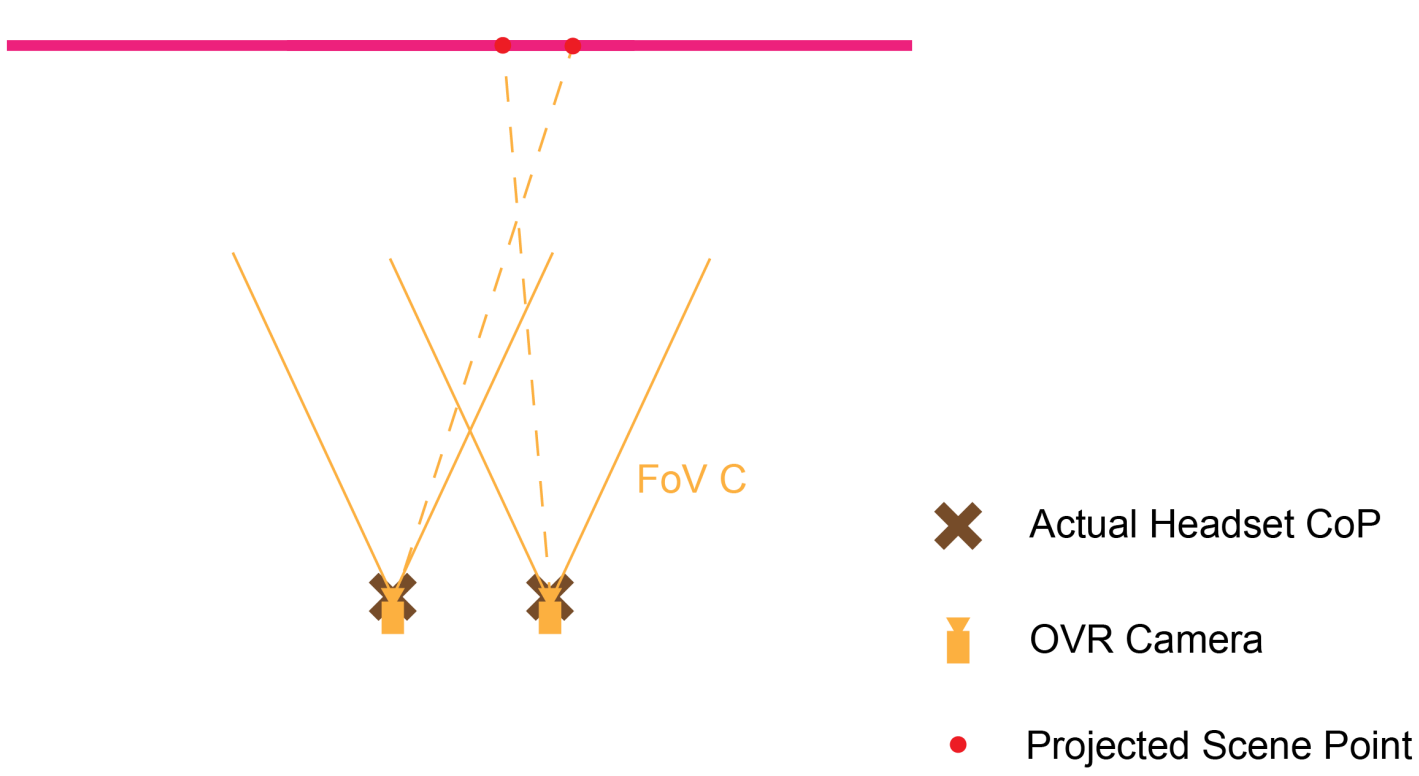

Fig. 17. The OVR camera subsequently views this quad and sends its frame buffer to the actual headset display.